\documentclass[twocolumn]{aastex631}
\usepackage{multirow}
\usepackage{xspace}
\usepackage{booktabs}
\usepackage{calligra} 

\usepackage{graphicx}

\usepackage{amsmath}

\newcommand{\Msun}{\,{\rm M_\odot}}
\newcommand{\fedd}{\,{f_{\rm Edd}}}
\newcommand{\Mbh}{M_\bullet}
\newcommand{\sag}{Sgr A$^\star$}

\newcommand{\HOLESOM}{\texttt{HOLESOM}}

\begin{document}

\title{HOLESOM: Constraining the Properties of Slowly-Accreting \\ Massive Black Holes with Self-Organizing Maps}

\author[0009-0002-5758-6025]{Valentina La Torre}
\affiliation{Center for Astrophysics $\vert$ Harvard \& Smithsonian, 60 Garden St, Cambridge, MA 02138, USA}
\affiliation{Department of Physics and Astronomy, Tufts University, Medford, MA 02144, USA}

\author[0000-0001-9879-7780]{Fabio Pacucci}
\affiliation{Center for Astrophysics $\vert$ Harvard \& Smithsonian, 60 Garden St, Cambridge, MA 02138, USA}
\affiliation{Black Hole Initiative, Harvard University, Cambridge, MA 02138, USA}

\begin{abstract}
Accreting massive black holes (MBHs, with $\Mbh > 10^3 \Msun$) are known for their panchromatic emission, spanning from radio to gamma rays. While MBHs accreting at significant fractions of their Eddington rate are readily detectable, those accreting at much lower rates in radiatively inefficient modes often go unnoticed, blending in with other astrophysical sources. This challenge is particularly relevant for gas-starved MBHs in external galaxies and those possibly wandering in the Milky Way. We present \HOLESOM\footnote{\texttt{HOLESOM} is publicly available at:\\ \url{https://github.com/valentinalatorre/holesom}}, a machine learning-powered tool based on the Self-Organizing Maps (SOMs) algorithm, specifically designed to identify slowly-accreting MBHs using sparse photometric data. Trained on a comprehensive set of $\sim 20,000$ spectral energy distributions (SEDs), \HOLESOM\ can (i) determine if the photometry of a source is consistent with slowly-accreting MBHs and (ii) estimate its black hole mass and Eddington ratio, including uncertainties.
We validate \HOLESOM\ through extensive tests on synthetic data and real-world cases, including Sagittarius A$^\star$ (Sgr A$^\star$), demonstrating its effectiveness in identifying slowly-accreting MBHs. Additionally, we derive analytical relations between radio and X-ray luminosities to further constrain physical parameters. The primary strength of \HOLESOM\ lies in its ability to accurately identify MBH candidates, which can then be targeted for follow-up photometric and spectroscopic observations. Fast and scalable, \HOLESOM\ offers a robust framework for automatically scanning large multi-wavelength datasets, making it a valuable tool for unveiling hidden MBH populations in the local Universe.
\end{abstract}


\section{Introduction} 
\label{sec:intro}
Actively accreting black holes are the archetypical panchromatic astrophysical objects. Their surrounding environment is the theater of complex physical processes within, e.g., their accretion disk \citep{Shakura_Sunyaev_1976}, corona \citep{Haardt_1991} and jet \citep{Blandford_Znajek_1977}, resulting in an electromagnetic spectrum that extends from the radio to gamma rays \citep{Elvis_1994, Shang_2011, Mullaney_2011}.

Multi-wavelength studies of Active Galactic Nuclei, or AGN, are powerful tools for studying massive black hole (MBH, $\Mbh > 10^3 \Msun$) growth at high-$z$ (e.g., \citealt{Volonteri_2005}), testing General Relativity (e.g., \citealt{Psaltis_2020}), and understanding the evolution of galaxies (e.g., \citealt{Ferrarese_Merritt_2000, Gebhardt_2000}). Available data have proliferated over the last 30 years thanks to dedicated all-sky and pointed surveys at many different wavelengths.

MBHs are detected at the center of most massive galaxies, and tight relations between their mass and their host's physical properties testify to their co-evolution \citep{Magorrian_1998, Ferrarese_Merritt_2000, Gebhardt_2000, Kormendy_Ho_2013, Reines_Volonteri_2015, Pacucci_2023_JWST}.
While occupying a tiny volume in the cores of their host galaxies, MBHs have an oversized impact on the evolution of the entire galaxy via their feedback (e.g., \citealt{DiMatteo_2005, Sijacki_2007, Schawinski_2007, Greene_2020}).
Despite different and, in principle, unrelated growth modes, galaxies and their SMBHs are locked in a cosmic dance of co-evolution.
A complete picture of the evolution of galaxies, the building blocks of our Universe, ultimately requires a complete understanding of black holes' growth and feedback processes.

The population of highly-accreting MBHs contained in AGN is well studied in the local Universe, with surveys (e.g., the Swift-BASS, \citealt{Ananna_2017, Ananna_2022, Koss_2022}) able to represent an unbiased picture of the AGN luminosity function, the black hole mass function (BHMF), and the Eddington ratio distribution function (ERDF), where the Eddington ratio is defined as:
\begin{equation}
    \fedd = \frac{\dot{\Mbh}}{\dot{M}_{\rm Edd}} \, ,
\end{equation}
where $\dot{M}_{\rm Edd}$ is the mass accretion rate at which a relativistic thin accretion disk \citep{Novikov_Thorne} would radiate at the Eddington luminosity (see, e.g., the discussion in \citealt{Pacucci_Narayan_2024}), and $\dot{M}$ is the mass accretion rate onto the MBH. In general, the radiated luminosity $L$ is related to the accretion rate $\dot{M}$ via $L = \epsilon \dot{M}c^2$, where $\epsilon$ is the matter-to-energy radiative efficiency. 
In the standard, thin-disk accretion scenario \citep{Shakura_Sunyaev_1976}, valid for $10^{-2} \lesssim  \fedd \lesssim 1$, the matter-to-energy radiative efficiency is $\epsilon \approx 10\%$ and depends on the black hole spin \citep{Bardeen_1970, Thorne_1974}.
Highly-accreting MBHs are also observed at high redshifts, with the farthest confirmed AGN now standing at $z = 10.6$ \citep{Maiolino_2023}, although the sample completeness is much lower in these cases.

On the contrary, even in the local Universe, the population of MBHs accreting at very low rates ($\fedd < 10^{-2}$) is sparsely studied.
For such low rates, the radiative efficiency $\epsilon$ decreases dramatically, possibly by orders of magnitude, as we enter the domain of advection-dominated accretion flows (ADAF, \citealt{Narayan_1994, Abramowicz_1995, Narayan_McClintock_2013}). 
As the conditions for large accretion rates are rare, especially at low redshift \citep{Power_2010}, most MBHs likely accrete in ADAF mode \citep{Narayan_2022}.
An example of such objects is the SMBH at the center of the Milky Way (MW): Sagittarius A$^\star$ (Sgr A$^\star$, in short). Recently imaged by the Event Horizon Telescope collaboration \citep{EHT_SgrA}, this SMBH is accreting at $\fedd \sim 10^{-7}$ (see, e.g., \citealt{Ressler_2018}) and was initially detected by studying the dynamics of the stars in its immediate vicinity \citep{Ghez_2000, Genzel_2000}.

Using various methodologies, spanning radio \citep{Reines_2020_radio}, optical \citep{Reines_2013_optical}, X-ray \citep{Baldassare_2015}, and coronal lines \citep{Molina_2021}, several samples of tens or even hundreds of MBHs in local dwarf galaxies exist, with typical distances up to $\sim 200$ Mpc. However, these surveys are biased towards intrinsically brighter MBHs, with an emission significantly brighter than other sources of stellar origin \citep{Pacucci_2021_active_fraction}.
For example, a $10^5 \Msun$ MBH has to radiate at $\sim 10\%$ of its Eddington luminosity to produce detectable broad H$\alpha$ that is discernible within $200$ Mpc \citep{Reines_2013_optical}. A $10^4 \Msun$ MBH would need to accrete at Eddington. These numbers are $10^6-10^7$ times higher than the Eddington ratio of Sgr A$^\star$. Thus, the population of MBHs with extremely low accretion rates, similar to that of Sgr A$^\star$, is, arguably, entirely unexplored.

Detecting electromagnetically MBHs accreting at $\fedd \ll 1$, even in the very local Universe, is challenging because these sources appear unremarkable in their field and can be easily confused with other sources of stellar origin or even with background, high-$z$ sources. Albeit at a smaller mass scale, a recent attempt to detect electromagnetically the emission from Gaia BH3, a very nearby stellar-mass black hole of $M_\bullet \approx 33 \, \rm M_\odot$, led only to an upper limit \citep{Cappelluti_2024}.

For these reasons, slowly accreting MBHs in local galaxies can be selected only via a massively multi-wavelength approach. Such objects can be discerned from other contaminant categories only by collecting numerous photometric points in different bands and comparing this photometry with ADAF SEDs generated either analytically (see, e.g., \citealt{Mahadevan_1997, Pesce_2021}), or with general-relativistic magneto-hydrodynamic (GRMHD) simulations of the accretion flow, down to a few gravitational radii (see, e.g., \citealt{grMonty_2009}). Advanced computational techniques can lead to better identification and characterization of such sources.

Machine learning (ML) techniques offer an exciting prospect for selecting and characterizing the physical properties of slowly accreting MBHs in the local Universe. Once an ML technique is trained with the most extensive available set of ADAF SEDs, it learns the patterns between observables and physical underlying parameters such that it can be deployed to analyze multi-wavelength data fields pertaining to the local Universe.
Such codes are easily scalable and can be fed massive amounts of data once proper source-matching techniques between catalogs are used (see, e.g., the software \texttt{LYRA}, \citealt{Peca_2021}).

In this study, we present \HOLESOM, a software tool based on the unsupervised ML technique of Self-Organizing Maps. This software represents the first attempt at ``black hole cartography'', aimed at identifying and mapping slowly-accreting MBHs in the correct location of the $\Mbh$ vs. $\fedd$ parameter space.
This study is organized as follows. In Sec.\,\ref{sec:data}, we present the analytical SEDs used to train our model. Then, Sec.\,\ref{sec:method} describes the ML methods we employ. In Sec.\,\ref{sec:results}, we present our results. We test our software with artificial (Sec.\,\ref{subsec:missing_data}) and real-world examples (Sec.\,\ref{subsec:SgrA}). Then, we show that \HOLESOM\ can automatically identify the photometry of slowly-accreting MBHs candidates while rejecting different astrophysical sources (Sec.\,\ref{subsec:exclusion_test}). Sec.\,\ref{sec:process} gives as overview of the procedure for utilizing \HOLESOM. We also present relationships between radio luminosity, X-ray luminosity, black hole mass, and Eddington rate (Sec.\,\ref{sec:relations}). We conclude with Sec.\,\ref{sec:conclusion} discussing future applications of \HOLESOM.

\begin{figure*}[ht!]
    \centering
    \includegraphics[width=0.48\textwidth]{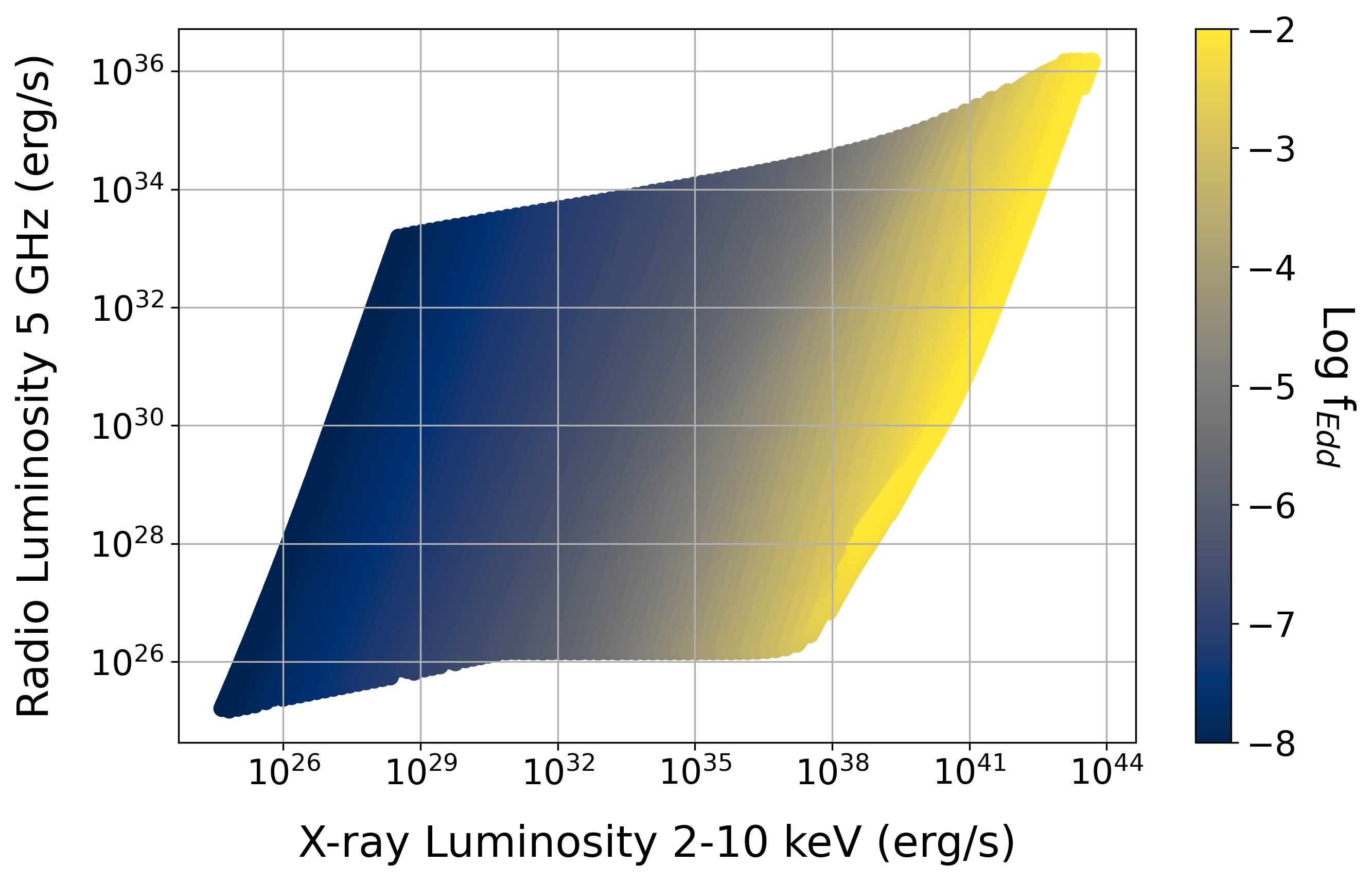}\hfill \includegraphics[width=0.48\textwidth]{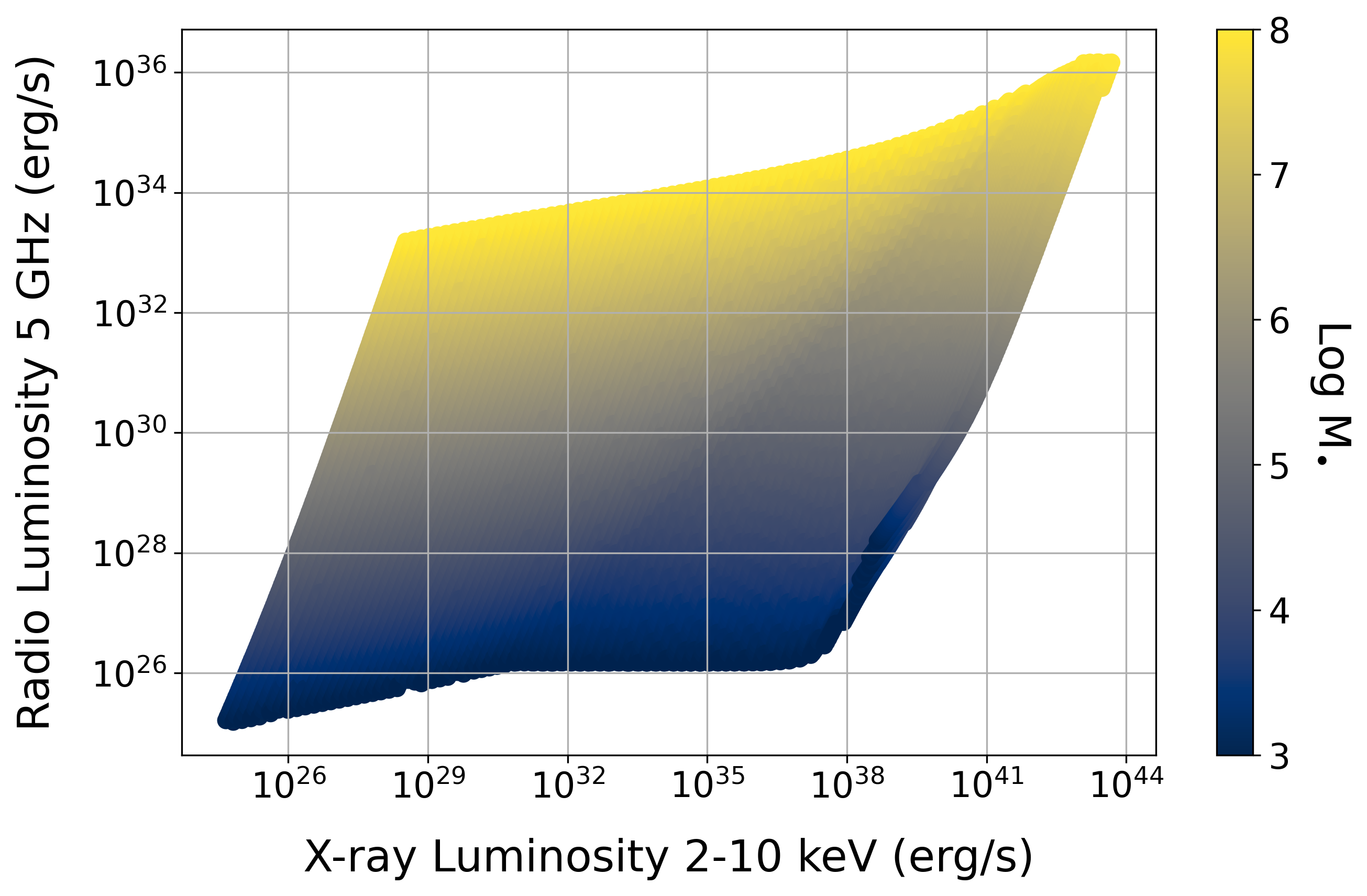}
    \caption{Luminosity at $5$ GHz versus the integrated X-ray luminosity at $2-10$ keV from our SED catalog, color-coded by $f_{Edd}$ (left panel) and $\Mbh$ (right panel). These plots demonstrate that, while there is no evident correlation between radio and X-ray luminosity, the latter generally increases with $\fedd$, while the former with $\Mbh$. 
    }
    \label{fig:Lradio_X}
\end{figure*}

\section{Spectral Energy Distributions}
\label{sec:data}
Training the ML-powered \HOLESOM\ to recognize and correctly classify the properties of MBHs requires a large library of ADAF SEDs. One method to obtain realistic ADAF SEDs is to execute a GRMHD simulation of a MBH of a given mass, and ray trace its spectral emission with codes such as \texttt{grMonty} \citep{grMonty_2009}. However, this method is computationally expensive if a number $N \gg 1$ of training SEDs are required. In fact, to obtain a reliable spectrum, about 100 different GRMHD snapshots are typically necessary \cite{grMonty_2009}, with each snapshot demanding approximately half an hour of computational time on a single core. For our ML application, the number of required training SEDs is typically $N \sim 10^4$. 

In order to obtain a large library of SEDs to develop \HOLESOM, we then rely on analytical models, firstly developed by \cite{Mahadevan_1997} and further developed and coded by \cite{Pesce_2021}. This model describes self-similar ADAF, characterizing the local properties of the accreting gas based on parameters such as mass, accretion rate, radius, and viscosity parameters, among other variables. In this regime, the accreting gas forms a two-temperature, optically thin plasma where ions are at their virial temperature while electrons remain cooler. The pressure arises from both gas and magnetic contributions. The emission mechanisms include synchrotron radiation, inverse Compton scattering, and bremsstrahlung \citep{Mahadevan_1997}.
This model calculates the spectral luminosity $L_\nu$ emitted by black holes of any mass, up to an Eddington ratio of $\fedd = 10^{-1.7}$; the only two parameters required to generate the SED are the black hole mass $\Mbh$ and the Eddington ratio $\fedd$. 

To construct our SED library, we compute $L_\nu$ at $54$ frequencies spanning from $5.0\times10^9$ to $1.9\times10^{19}$ Hz, namely from $5$ GHz in the radio band to $10$ keV in the X-ray. We generated SEDs for MBHs with mass and Eddington ratio in the ranges $\log_{10}(\Mbh/\Msun)$ = [3,8] and $\log_{10}\fedd$ = [-8,-2], respectively. Each range was divided into $140$ logarithmic bins, thus obtaining a total of $140^2 = 19600$ spectra. From this data set, we removed a small set of outliers (i.e., $\sim 0.3\%$ of the initial distribution) because the resulting SED showed clear signs of numerical problems, which occurred typically for extremely low values of the Eddington ratio $\fedd \sim 10^{-8}$. Hence, the final number of SEDs used to train our ML algorithm is $19535$.

Figure \ref{fig:Lradio_X} displays the general distribution of $\nu L_\nu$ in the radio ($5$ GHz) vs. the integrated X-ray luminosity ($2-10$ keV) for our SEDs. No strong correlation between radio and X-ray luminosity is evident for such a vast collection of black hole masses and Eddington ratios. However, the left panel, color-coded by $\log_{10}\fedd$, shows that the X-ray luminosity increases as the Eddington ratio increases. In contrast, the right panel, color-coded by $\log_{10}\Mbh$, shows that radio luminosity increases with black hole mass.

\section{Methods}
\label{sec:method}
In this Section, we first describe the ML algorithm underlying \HOLESOM\ and its training process, followed by the methods employed to derive analytical equations in Sec.\ref{sec:relations}. 

\subsection{Self-Organizing Maps}
\label{subsec:SOM}
We develop \HOLESOM\ on the basis of the ML algorithm named Self-Organizing Maps (SOMs) \citep{kohonen1982}, and make it publicly available at \href{https://github.com/valentinalatorre/holesom}{https://github.com/valentinalatorre/holesom}. \\
SOMs are a type of unsupervised artificial neural network (see, e.g., \citealt{ML_2019} for a review) that reduces a high-dimensional parameter space to a lower-dimensional representation while preserving the original topology. In this way, similar objects in the high-dimensional parameter space will be grouped in the final map, maintaining the intrinsic structure of the input data. In this study, we reduce the dimensionality to $2$, as this choice facilitates visualization. We use the publicly-available code \textsc{SomPY}\footnote{\url{https://github.com/sevamoo/SOMPY}} library \citep{sompy} to construct and train our SOM. 

Each pixel of the 2D map is characterized by a weight vector corresponding to a position in the higher-dimensional parameter space of the input data, normalized to the unit variance with a mean of zero. The weight vectors are initialized through principal component analysis (PCA; \citealt{PCA}) and then updated iteratively to minimize the Euclidean distance to the input data and reduce the quantization error. For each input data point, the cell whose weight vector has the smallest Euclidean distance to it is identified as the Best Match Unit (BMU).

We adopt an unsupervised approach with SOM rather than a supervised method for several reasons. First, supervised models require extensive hyperparameter tuning, model parameter and optimization choices that significantly impact model performance and computational time. In contrast, SOM operates directly in the data space with a simpler set of hyperparameters and does not require strong prior assumptions about model complexity. Second, we desire uncertainty in our results rather than a point estimate.
Obtaining uncertainty estimates with supervised methods typically requires assuming an underlying data distribution (e.g. Gaussian, multimodal). SOM is non-parametric and can be used to construct a posterior probability distribution function (PDF) entirely from the data itself \citep{Lopez-Rubio_2010, Hemmati_2019, LaTorre_2024}. Third, SOM provides an intuitive 2D representation of the data, enabling the identification of relationships and clusters that might not be apparent in a supervised framework. Finally, with SOM, we can easily implement methods for handling missing data without retraining the model \citep{LaTorre_2024}, as discussed in Section \ref{subsec:missing_data}. This is crucial when dealing with sparse photometric observations, which can vary across test sources, as is expected for MBH data. In summary, SOM can provide visualization (Sec.\,\ref{subsec:training}), parameter estimation with PDFs (Sec.\,\ref{subsec:missing_data}, \ref{subsec:SgrA}), and classification (Sec.\,\ref{subsec:exclusion_test}) in a single framework, offering a more interpretable and flexible model with lower computational complexity.

\begin{figure*}
    \centering
    \includegraphics[width=0.6\textwidth]{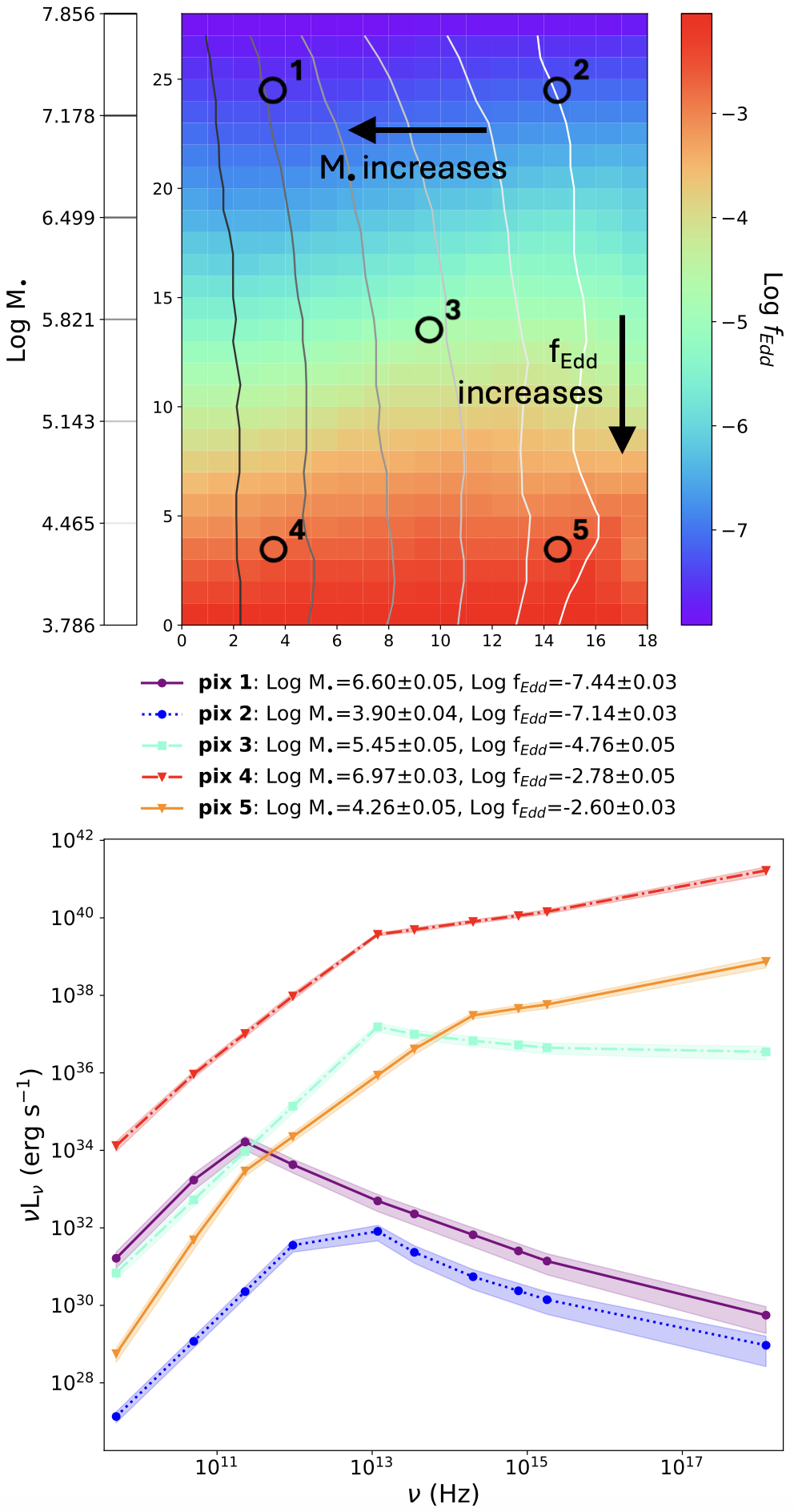} 
\caption{\textit{Top panel}: Trained SOM color-coded by $\log_{10}\fedd$. The different shades of gray contours indicate $\log_{10}\Mbh$ values, as shown in the left color bar. The open circles represent five pixels (pixel 1 = [3,24], pixel 2 = [14,24], pixel 3 = [9,13], pixel 4 = [3,3], pixel 5 = [14,3]), covering different combinations of $\Mbh$ and $\fedd$. From these pixels, we select test MBHs for the analysis of Sec. \ref{sec:results}. \textit{Bottom panel}: SEDs of MBHs located in the five test pixels. The shaded regions indicate the standard deviation of the spectral luminosity in that pixel, whose median values for $\log_{10}\Mbh$ (in units of $\Msun$) and $\log_{10}\fedd$ are reported in the legend with the corresponding standard deviation of their distribution in their pixel.}
\label{fig:som_test_sed}
\end{figure*}

\subsubsection{Training and Visualization} \label{subsec:training}
In SOM algorithms, input features are individual attributes of the dataset that are fed into the SOM for analysis. These features define the higher-dimensional space, which the SOM reduces to the 2D grid. The input features of \HOLESOM, which are given by different photometric data points for a given source, are summarized in Table \ref{tab:input_features}. 

These input features were carefully selected based on two main criteria. First, we aimed to balance homogeneous spectral coverage with practical constraints of ML applications and observational limitations. While a continuous sampling approach is theoretically ideal, it is impractical for MBH applications, where continuous spectral coverage is rarely available. Indeed, in reality, MBHs have only a few detections, and adding more features would only exacerbate the issue of missing data, as ML models require consistency between training and test sets. Therefore, we select a minimal yet representative set of features that ensures homogeneous coverage across the SED while balancing the constraints of ML algorithms and the observational limitations of MBHs. By adopting a representative and widely spaced sampling of the SED, we mitigate the impact of missing data--an expected issue with real data, though we do not know a priori which bands will be missing for a given source--while preserving meaningful physical information. The treatment of missing data is further discussed in Section \ref{subsec:missing_data}. 
\begin{table}[h]
\centering
\caption{Summary of input features for the SOM algorithm.}
\begin{tabular}{ll}
\toprule
\textbf{Input Feature} & \textbf{Frequency (Hz)} \\
\midrule
\midrule
\textbf{Radio Luminosity} \\
\quad 5 GHz  & $5 \times 10^9$ \\
\quad 50 GHz & $5 \times 10^{10}$ \\
\quad 230 GHz & $2.3 \times 10^{11}$ \\
\midrule
\textbf{X-ray Luminosity} \\
\quad 2-10 keV & $4.8 \times 10^{17} - 2.4 \times 10^{18}$ \\
\midrule
\textbf{SED Slopes} \\
\quad Radio (5-950 GHz) & $5 \times 10^9 - 9.5 \times 10^{11}$ \\
\quad Far-Infrared (316 - 25.5 $\mu$m) & $9.5 \times 10^{11} - 1.2 \times 10^{13}$ \\
\quad Mid-Infrared (25.5 - 8.6 $\mu$m) & $1.2 \times 10^{13} - 3.5 \times 10^{13}$ \\
\quad Near-Infrared (8.6 - 1.5 $\mu$m) & $3.5 \times 10^{13} - 2 \times 10^{14}$ \\
\quad Optical (1.5 - 0.4 $\mu$m) & $2 \times 10^{14} - 7.7 \times 10^{14}$ \\
\quad UV (0.4 - 0.17 $\mu$m) & $7.7 \times 10^{14} - 1.8 \times 10^{15}$ \\
\bottomrule
\end{tabular}
\label{tab:input_features}
\end{table}
Second, we prioritize features most sensitive to physical parameters such as $\Mbh$ and $\fedd$. To identify these, we computed the distance correlation between $\Mbh$ and $\fedd$ with all the luminosities $\nu L_\nu$ across 54 frequencies from radio to X-ray, as well as all possible two-frequency combinations of SED slopes (1,326 combinations). After selecting the features most strongly correlated with these parameters, we trained multiple SOMs with different feature sets. The final 10 features reported in Table \ref{tab:input_features} led to the smallest dispersions in $\Mbh$ and $\fedd$ within SOM cells. 

The size and geometry of our SOM are optimized for data clustering and performance, following the method described in \cite{D19}, resulting in a final map of $28\times18$ pixels. Black holes with similar SEDs are placed in the same or nearby pixels on the map. Since a given SED depends on its Eddington ratio and black hole mass, objects in the same pixel share similar SEDs (and, hence, similar $\fedd$ and $\Mbh$). 

We compute the median values of $\fedd$ and $\Mbh$ for each pixel. The top panel of Fig. \ref{fig:som_test_sed} shows the fully trained SOM color-coded by median $\log_{10}\fedd$, with values smoothly increasing vertically from top to bottom. In contrast, median $\log_{10}\Mbh$ values progressively increase leftwards, as shown by the shades of gray contours with values indicated in the color bar on the left. 

The pixel distributions of $\Mbh$ and $\fedd$ are generally narrow, with $\sigma_{\rm STD}$ ranging between $0.03-0.09$ dex for $\log_{10}\Mbh$ and $0.02-0.08$ dex for $\log_{10}\fedd$. To demonstrate that the clustering process was successful, the bottom panel of Fig. \ref{fig:som_test_sed} shows five median SEDs, along with their standard deviations, from five test pixels circled on the map in the top panel. These pixels represent different regimes of $\Mbh$ and $\fedd$: (1) is high $\Mbh$ and low $\fedd$, (2) is low $\Mbh$ and low $\fedd$, (3) is intermediate $\Mbh$ and intermediate $\fedd$, (4) is high $\Mbh$ and high $\fedd$, (5) is low $\Mbh$ and high $\fedd$.
The shaded regions indicate that the SEDs in each pixel are very well-constrained, with small deviations around the median.

\subsection{Polynomial and Symbolic Regressions}
\label{subsec:regression}
In this Section, we describe the two regression techniques employed to derive the analytical relations presented in Sec.\,\ref{sec:relations}: polynomial and symbolic regression. 

We first employ polynomial regression to capture the relationships between $\Mbh$ and the radio luminosity at 5 GHz, and between $\fedd$ and the X-ray luminosity at 2-10 KeV. Polynomial regression fits the data to a polynomial function of a given degree, allowing for the modeling of non-linear trends without increasing the complexity of the model. We used the \texttt{numpy.polyfit} function from the Numpy library for this task and chose a second-degree polynomial fit. This choice was based on the observation that higher-degree polynomials did not significantly improve the fit, and the data distributions did not exhibit complex behaviors. 

However, we employ symbolic regression for more complex relationships where a polynomial fit would be insufficient. This machine-learning technique finds mathematical equations to describe data by combining basic operators like addition and multiplication. Unlike polynomial regression, which fits data to a predefined function, symbolic regression explores a vast space of possible equations to identify the one that best captures the relationships in the data. We used the \textsc{PySR}\footnote{A description of PySR is available at this \href{https://github.com/MilesCranmer/PySR}{link}.} library \cite{Cranmer_2023}, which employs an evolutionary algorithm inspired by natural selection. The algorithm starts with a group of randomly generated equations (i.e., the ``individuals'') and gradually improves them. Each step selects the best-performing equations based on how well they fit the data, applies random changes (the ``mutations'') to generate new equations, and replaces weaker ones with stronger ones. This process repeats many times, evolving the equations toward better solutions. 

For our training, we employ a representative subset of 1000 randomly selected objects and configured the model with \texttt{populations}=100, allowing for the exploration of a diverse set of solutions by evolving multiple independent groups of candidate equations. We set \texttt{niterations}=50 to allow enough iterations for each population to evolve through mutation and selection. Additionally, we allow basic binary operators \{+,-,$\times$,/\} and unary operators \{square, cube, exp\} during the symbolic regression.

\section{Results}
\label{sec:results}

In this Section, we apply the ML method described thus far, adopting a progressive approach. 
First, we test how well we can constrain the main physical parameters of a putative source if data is available only in a very limited set of frequencies (Sec.\,\ref{subsec:missing_data}). This, of course, will be the case with any real-world application. Second, we test \HOLESOM\ with the archetypal example of an ADAF MBH: \sag\ (Sec.\,\ref{subsec:SgrA}). In doing this, we highlight why this source is peculiar and possibly not representative of the majority of sources to which our methodology can be applied. Then, in Sec.\,\ref{subsec:exclusion_test}, we show that \HOLESOM\ can effectively identify MBHs accreting in ADAF mode while filtering out sources with non-MBH photometry.

\subsection{Evaluating $\fedd$ and $\Mbh$ ranges with limited data}
\label{subsec:missing_data}

We now test the fully-trained SOM to investigate its performance in real-world scenarios. In an ideal situation, a given source possesses a large wealth of multi-wavelength data, spanning from radio to X-ray. In most cases, however, only a limited amount of data can be securely associated with a given source, with several data points typically lower than the $10$ input features with which our code was trained. The relevant question is: How well can the physical parameters of the MBH be constrained if one or more multi-wavelength data points are missing? 

At this stage, it is important to understand that \HOLESOM\ is a tool whose main goals are to identify slowly-accreting MBHs candidates and provide an \textit{estimate} of their physical parameters, based on very limited photometric data. Once candidates are identified, additional follow-up photometric data will improve our estimate, while spectroscopy will ultimately investigate their nature in detail.

Hence, our goal is to investigate whether, with scarce data, we can: (i) correctly classify the mass category of a MBH and (ii) constrain its Eddington ratio within the upper end of the ADAF range (i.e., $\fedd \sim 10^{-2} - 10^{-3}$), or at severely low rates (i.e., $\fedd \ll 10^{-3}$). We first discuss the ideal scenario with no missing observations. Next, we explain how to estimate parameters when data is missing, beginning with the case of a single missing photometric point and concluding with scenarios of multiple missing observations. 

\begin{figure*}
    \centering
    \hspace*{-2cm}
   
     \includegraphics[width=1.0\textwidth, trim={1cm 0 0 0}]{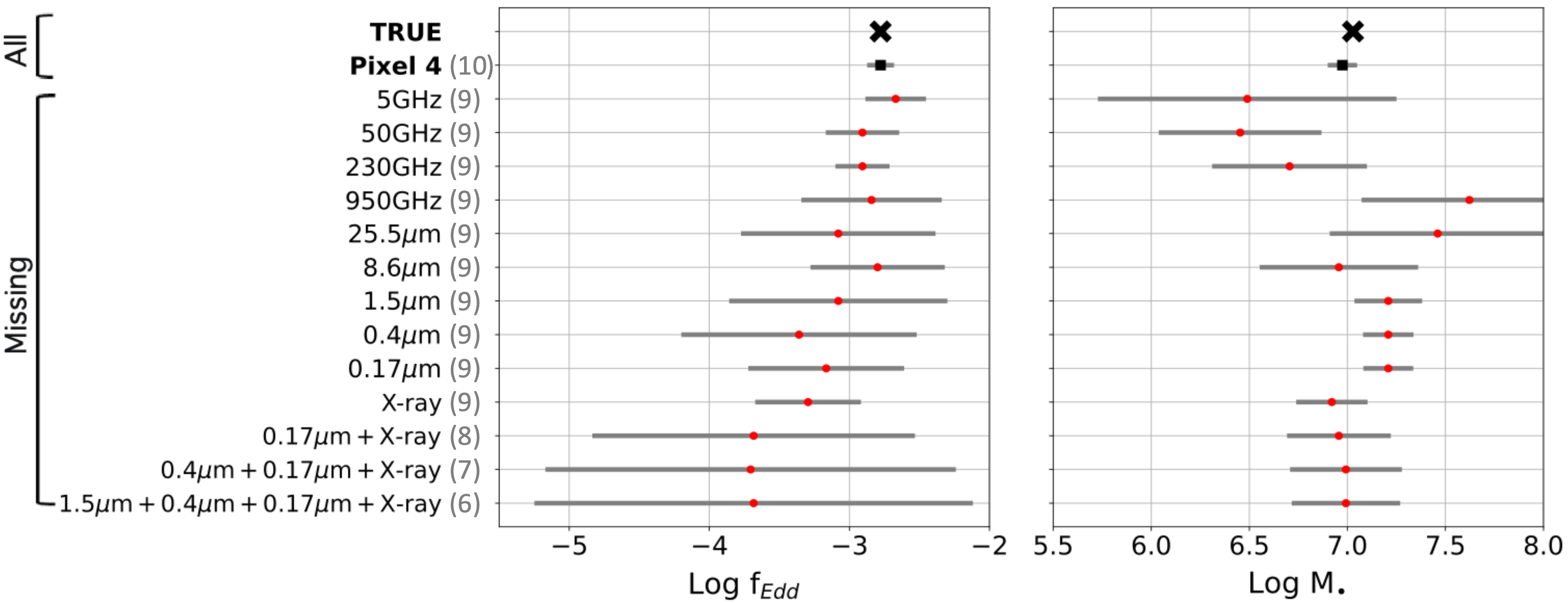} 
\caption{For a test MBH located in \textit{pixel 4}, we illustrate the values of Log $\fedd$ (left) and $\Mbh$ (right) estimated with \HOLESOM, in various cases of data availability. The numbers in parentheses indicate how many of the 10 bands are available. In the top rows, labeled `All', we show the true values of the test MBH, marked by a black cross. The black square indicates, when all bands are available, the median of the parameter distribution within \textit{pixel 4}, with the grey line showing the standard deviation of that distribution. In the lower rows, labeled `Missing', we show the parameter ranges after recovering one or multiple missing observations (Section \ref{subsec:missing_data}). In these cases, the red circle marks the peak of the probability distribution function for the estimated parameter after recovering missing data, while the grey line again represents the standard deviation. Additional test MBHs from other pixels are shown in the Appendix\,\ref{app:missing_data}.}
\label{fig:result_example}
\end{figure*}

We take one test MBH from each of the five test pixels described in Sec. \ref{subsec:SOM}. When all input features are available, we assign the median values of $\fedd$ and $\Mbh$ from the corresponding pixel to the test MBH. The uncertainty is given by the standard deviation of these parameter distributions within the pixel. The accuracy of this approach is illustrated in Fig.\,\ref{fig:result_example} for ``Pixel 4", showing that the estimated parameters are very accurate when no photometry information is missing.

For a more realistic scenario, we remove one photometric band at a time. We then recover that specific missing information following the method detailed in \cite{LaTorre_2024}, to which the interested reader is referred. In summary, this method consists of randomly drawing $5000$ new values of the missing feature from the distribution of the rest of the training sample. This approach is justified as we can reasonably assume that the missing value, due to non-detection, falls within the same distribution as the training sample, which contains values for the MBH population, and it is not an outlier or a peculiar case. Then, given the new set of input features, we project the MBH onto the SOM, creating a 2D likelihood surface of the possible pixel occupied by the MBH. From that, we extract the probability distribution function of $\fedd$ and $\Mbh$ for that MBH, given the recovered missing data.

As an example, Figure \ref{fig:result_example} shows the ranges of $\fedd$ (left panel) and $\Mbh$ (right panel) for recovered missing data of a test MBH selected from the test pixel 4 (see Figure \ref{fig:som_test_sed}). Results for the other test pixels are reported in Appendix \ref{app:missing_data}.
Here, we begin by discussing the process of constraining $\fedd$ and then proceed to $\Mbh$, starting with a single missing observation and then addressing multiple missing observations. 
For $\fedd$, we observe that the ranges are consistent with the true value, especially in the upper sub-Eddington regime around 10$^{-3}$-10$^{-2}$ (i.e., Figure \ref{fig:results1} in the Appendix), where $\fedd$ is mainly affected by IR, optical, UV, and X-ray observations. This result also holds in the low end of $\fedd$ around $10^{-8}-10^{-7}\Msun yr^{-1}$ (i.e., the top two panels in Figure \ref{fig:results} in the Appendix). Even with missing data, we can constrain $\fedd$ to be in the upper or lower ranges in the sub-Eddington regime.
This result slightly differs for a MBH in an intermediate sub-Eddington regime (i.e., $\sim10^{-5}$), as shown in the bottom row of Figure \ref{fig:results}, for a MBH in pixel 3. Here, radio observations start to constrain $\fedd$; specifically, recovering the missing luminosity at $950$ GHz would place this object in a deeper sub-Eddington regime. 

Recovering missing observations yields consistent ranges of $\Mbh$ with the true values for cases at the high end of the black hole mass (i.e., $\sim10^7 \Msun$). Unlike $\fedd$, radio and MIR luminosities mainly influence the mass range. This pattern also holds in the opposite regime, i.e., at the low mass end (pixel 2 and 5 in Figure \ref{fig:results}), where recovering missing observations still yields consistent mass ranges compared to the true value. However, missing luminosity at $5$ GHz and/or $8.6 \, \mu$m could result in mass regimes differing by $\sim 1-2$ orders of magnitude. For the intermediate mass case (pixel 3), the mass ranges are consistent with the true value within the margin of error.

While the previous results focused on recovering a single missing observation, we now examine the results when multiple missing observations are recovered. The bottom rows of Figure \ref{fig:result_example} show the ranges of $\fedd$ and $\Mbh$ for a test MBH from pixel 4 when several observations (from X-ray to IR) are missing. For results from the other test pixels, refer to Appendix \ref{app:missing_data}. The worst case among the five test MBHs is represented by black holes with very low $\fedd$ (i.e., $<10^{-7}$; see pixel 1 and 2). 
Conversely, for larger $\fedd$, this combination of missing observations produces $\fedd$ ranges consistent with the true value within the margin of error, even though these ranges can be broad. In essence, constraining $\fedd$ within a narrow range is challenging when X-ray, UV, optical, and IR are all missing. However, in this scenario, $\Mbh$ ranges remain consistent across all five test black holes (Figure \ref{fig:mass_more_missing}). As previously mentioned, $\Mbh$ estimation is not significantly affected by X-ray, UV, optical, and IR observations. 

In conclusion, in the ideal scenario of no missing data (i.e., $\sim 10$ photometric data points), $\fedd$ and $\Mbh$ are estimated with high accuracy. In case of missing X-ray data or, more generally, a single-band observation, \HOLESOM\ remains effective in correctly classifying MHBs within reasonable mass and accreting ADAF ranges. The uncertainty is higher in the case of numerous missing data points, but \HOLESOM\ remains effective in correctly identifying a slowly-accreting MBH and approximating its mass and Eddington ratio.

\subsection{The Golden Test: Sgr A$^\star$}
\label{subsec:SgrA}
\begin{figure*}
    \centering
    \includegraphics[width=1\textwidth]{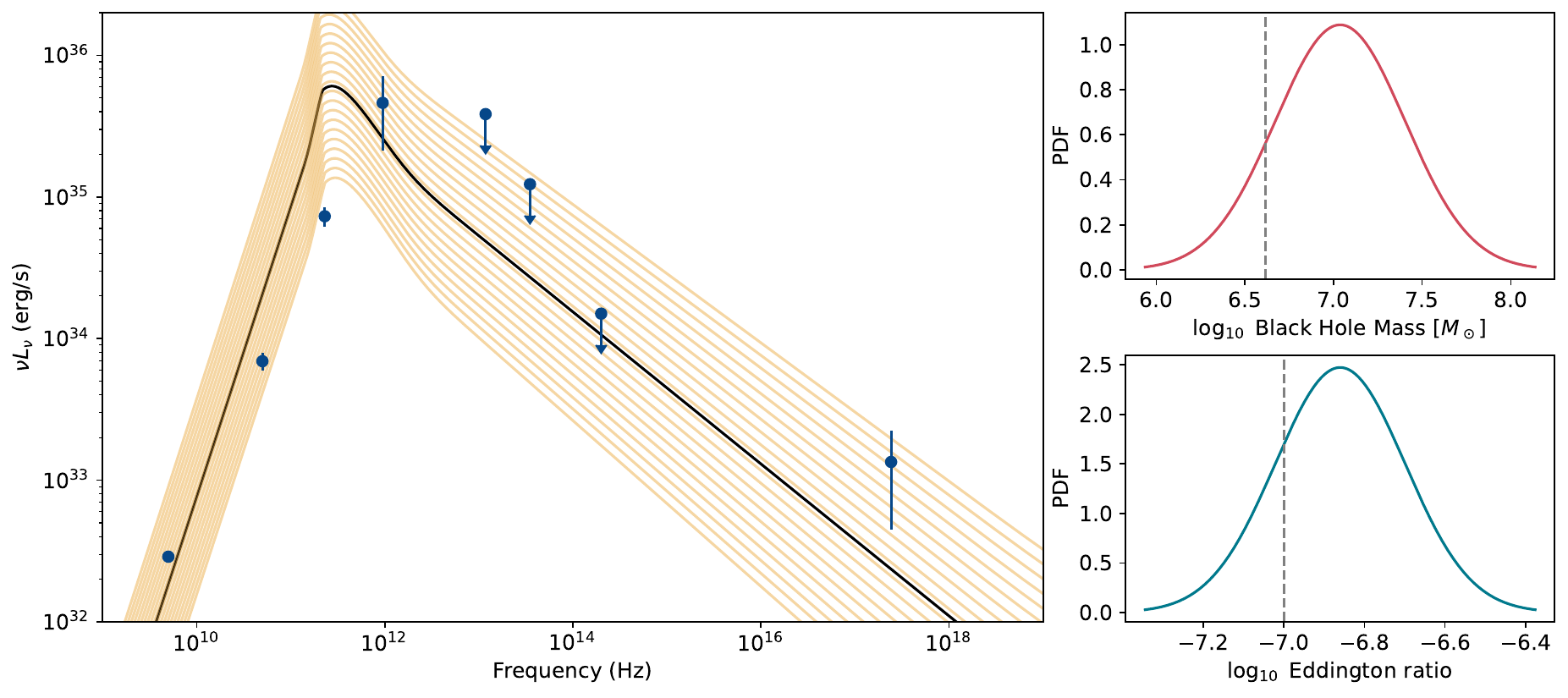}
    \caption{Test case for \sag. The left panel shows the SEDs selected by our code as the ones that best describe the data, shown in blue with their $1\sigma$ error bars. The black line shows our median SED, while the yellow region displays its $3\sigma$ uncertainty. For this test, we use only data from the radio, sub-millimeters, far infrared, and X-ray bands. The right panels show the inferred distributions for the black hole mass (top) and Eddington ratio (bottom), with their true values indicated by dashed lines.}
    \label{fig:sgrA}
\end{figure*}

We test \HOLESOM\ on the archetypal case of a MBH accreting in ADAF mode: \sag.
Although this MBH represents a perfect test case for our approach, we highlight two significant differences between \sag and the typical MBH that we will target with our code.
First, the mass of \sag, and, to some extent, its Eddington ratio, are known from a variety of independent observations, starting from the dynamics of stars in its immediate vicinity (see, e.g., \citealt{Ghez_2008}) to recent measurements by the Event Horizon Telescope \citep{EHT_SgrA}. The mass is $\Mbh = 4.0^{+1.1}_{-0.6} \times 10^6 \Msun$, with an Eddington ratio of $\sim 10^{-7}$. Our methodology, based on only a few photometric data points, aims at broadly characterizing the mass and accretion regime of the MBH, and cannot compete with such accurate measurements. However, it is instructive to check whether our estimate is compatible with the true value for \sag. 
Second, \sag is located in the Galactic center and characterized by heavy obscuration by interstellar dust (see, e.g., \citealt{Fritz_2016}), making optical and UV observations practically impossible. Targeting extragalactic MBHs, or off-centered wandering MBHs \citep{DiMatteo_2023_vast} in the Milky Way, will avoid this issue. Therefore, for this test, we fill-in optical and UV data before projecting \sag  on the trained map.

Radio observations at 5, 50, and 950 GHz are taken from \cite{Markoff_2001, Yuan_2003, Dodds-Eden_2009}; at 230 GHz from \cite{Bower_2019}; IR observations are taken from \cite{Melia_2001, Genzel_2003, Schodel_2007}, and the X-ray luminosity of \sag in its quiescent state is from \cite{Baganoff_2003}. The input features used are the same as those in Table\,\ref{tab:input_features}, except that we use only the SED slope between 5 and 950 GHz and use a new slope between radio (950 GHz) and X-ray. This additional slope helps track the synchrotron peak, as no information is employed for wavelengths in between. We also include photometric errors and upper limits following the approach described in \citep{LaTorre_2024}. We generate 50 new input datasets by drawing new random luminosities from a Gaussian distribution centered on the true luminosity with a width equal to the error. In the case of upper limits, we retain only values lower than the true luminosity. Each of these 50 datasets is then projected onto the trained SOM. As described in Section\,\ref{sec:results}, the ranges of $\Mbh$ and $\fedd$ are determined from the 2D likelihood surface of these datasets on the SOM. Figure\,\ref{fig:sgrA} shows the probability distribution function (PDF) of the derived $\Mbh$ and $\fedd$, compared to the true values for \sag, and the corresponding SED models, which are consistent with the observed photometric points and their errors.

\subsection{Identifying MBHs}
\label{subsec:exclusion_test}

\begin{figure*}
    \centering
    \includegraphics[width=0.52\linewidth]{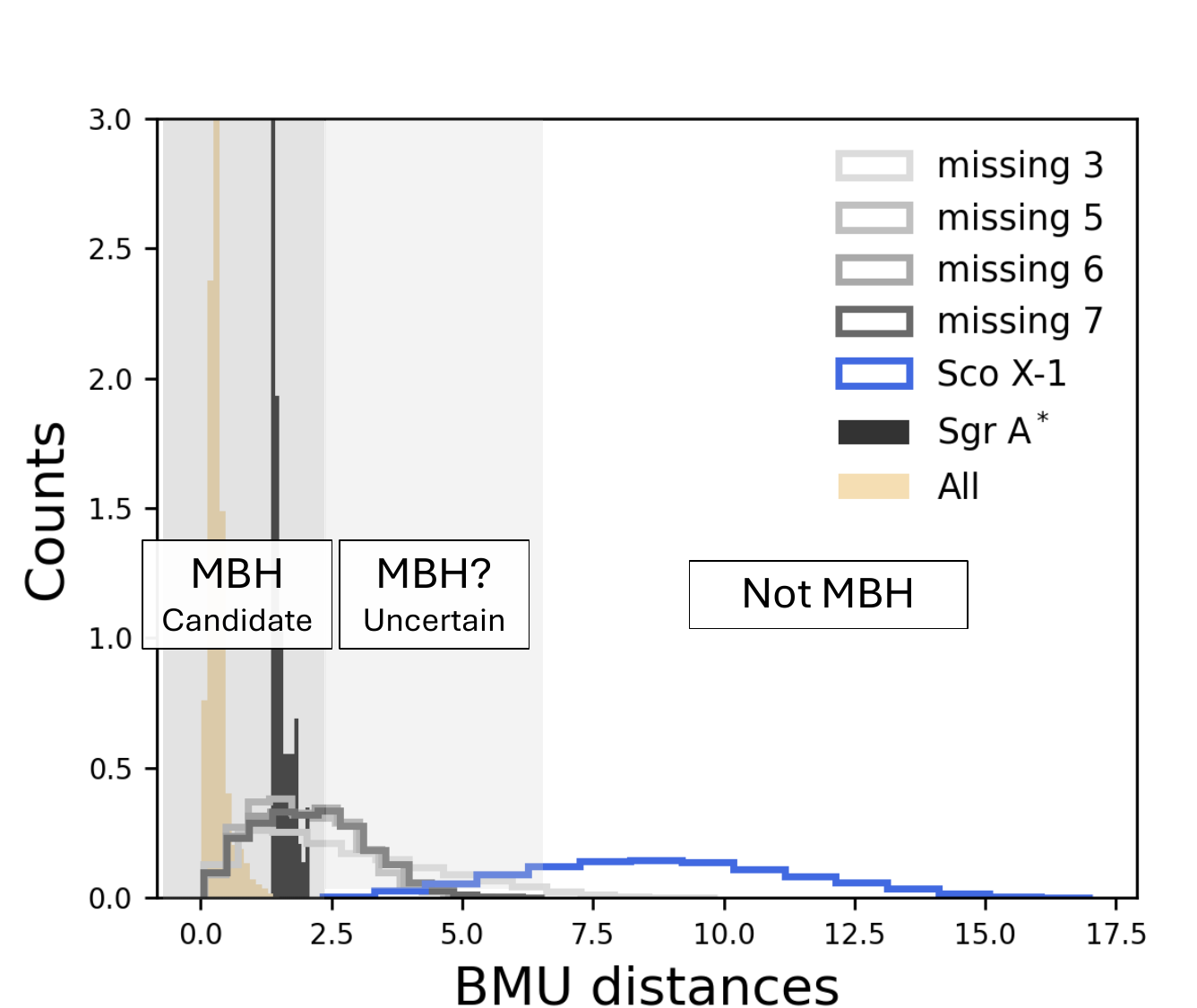}
    \includegraphics[width=0.45\textwidth]{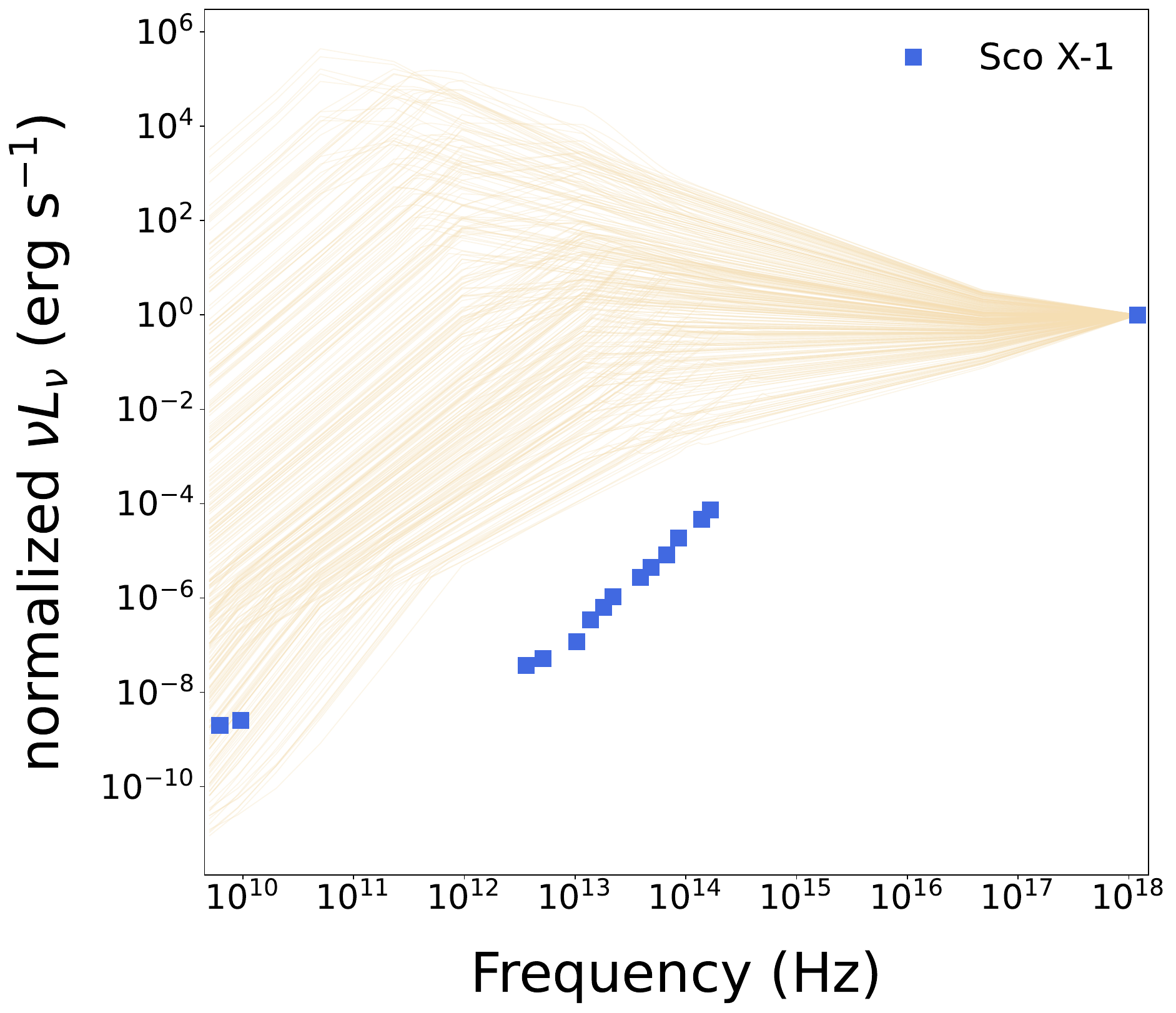} 
\caption{\textit{Left}: Sources in a catalog are identified as MBH candidates (or not) based on their BMU distance. BMU distance distributions are shown for MBHs with a complete feature set (orange), MBHs with recovered 3, 5, 6, and 7 recovered missing features (shades of gray), \sag (black), and Sco X-1 (blue). See Sec. \ref{subsec:exclusion_test} for details. \textit{Right}: Comparison between 300 models for MBH SEDs (orange lines), randomly drawn from the training set to cover all the physical parameter space, and the photometry for neutron star Sco X-1 (blue squares). Luminosities are normalized to the X-ray value.}
\label{fig:ScoX1}
\end{figure*}

With the new generation of telescopes increasing the volume of observational data, it is crucial to develop automated methods for systematically identifying MBH candidates before characterizing their properties and for follow-up observations. Currently, no standard automated approach exists for this task.

\HOLESOM\ is the first attempt to systematically identify slowly-accreting MBH candidates from a multi-wavelength catalog, leveraging its training on MBH SEDs. Since ADAF MBHs are the only population the algorithm has been trained on, when new sources are projected on the map, \HOLESOM\ effectively acts as a ``filter", retaining sources that resemble MBH templates while rejecting those that do not. Specifically, \HOLESOM\ can: i) \textit{recognize} objects with photometry similar to those in the training set by assigning them to SOM cells that best represent their observed SEDs, and ii) \textit{reject} any source whose observed photometry differs significantly from the MBH training set, classifying it as a non-MBH (or a MBH accreting at substantial levels).

When a test object is projected onto \HOLESOM, its identification is determined by the BMU distance (Sec.\,\ref{subsec:SOM}), which quantifies the Euclidean distance between the object's observed data and the closest SOM cell (the BMU). Each BMU represents an MBH template. We compute BMU distances for 19,000 MBH models with complete input features, 2,000 MBH models with recovered missing features, and two real cases--an ADAF MBH and a non-MBH source. Because we account for photometric errors and missing data, this results in a distribution of BMU distances rather than a single value.  
As illustrated in Figure \ref{fig:ScoX1}, BMU distance quantifies similarity to the MBH models making up the trained grid, allowing us to set identification thresholds: 
\begin{itemize}
    \item BMU distance $<$ 2.5: The test object is confidently identified as an MBH.
    \item BMU distance 2.5-7: The identification is uncertain, likely due to insufficient photometric coverage. Additional observations at currently missing wavelengths will refine classification. 
    \item BMU distance $>$ 7: The object is confidently classified as non-MBH, indicating it belongs to a different astrophysical population. 
\end{itemize}

A challenge arises when test objects lack complete photometric coverage. In the previous section, we fill in missing values using luminosity or SED slope distributions from the MBH training set. However, such distributions are not available for non-MBH objects. Applying MBH-based priors to non-MBH objects would make their SEDs appear more MBH-like, potentially blurring the distinction between the populations and reducing classification accuracy. Despite this challenge, we test \HOLESOM's identification capability with real-world examples and find that the method remains effective.

To evaluate \HOLESOM's performance, we project Sco X-1, a stellar-mass compact object, onto the map. As shown in Figure \ref{fig:ScoX1} (right), Sco X-1 has a remarkably different SED from those of MBHs accreting in ADAF mode. Of the 10 input features used for training (Table \ref{tab:input_features}), six are missing. We only have the radio luminosity at 5 GHz, the X-ray luminosity, and the IR luminosities at 25.5, 8.6, and 1.5 $\mu$m, i.e., five bands out of the 10 needed to construct the full feature set. Despite recovering missing data using MBH priors (which could make Sco X-1's SED more similar to MBHs), Sco X-1 still exhibits a high BMU distance $\gtrsim$ 7 (Figure \ref{fig:ScoX1} left). Note that MBHs with even more missing features still maintain lower BMU distances (0-5). This confirms that \HOLESOM\ correctly rejects Sco X-1 as an MBH candidate, placing it at the edge of the map (i.e., in the SOM location [18,0]), where outliers are typically located (see, e.g., \citealt{Davidzon_2019}). Conversely, when we project \sag onto the map (Sec.\,\ref{subsec:SgrA}), its BMU distance remains below 1.8 (Figure \ref{fig:ScoX1} left), reinforcing its identification as an ADAF MBH with high confidence, even with missing data.

By implementing the BMU distance threshold, \HOLESOM\ provides a robust, automated framework for identifying ADAF MBH candidates in large datasets. Even when dealing with missing data, the method remains effective, ensuring reliable discrimination between MBH candidates and other astrophysical populations.

\section{How to use \HOLESOM\ with new observations}\label{sec:process}
The population of slowly accreting MBHs remains elusive due to the challenge of detecting them. Only a few candidates have been observed, often with sparse multi-wavelength data \citep[e.g.,][]{Nguyen_2019, Reines_2022Nat}
. Moreover, even knowing where to look for such objects is difficult. This work introduces \HOLESOM\ as a tool to identify slowly accreting MBH candidates in large multi-wavelength catalogs containing diverse astrophysical sources. \HOLESOM\ can also characterize their physical properties $\Mbh$ and $\fedd$. Once identified, these candidates can benefit from follow-up observations to refine their parameters and spectroscopy to investigate their nature in detail. While this paper presents and validates the methodology using specific examples (\sag and Sco X-1), a large-scale application to compile a catalog and project it onto \HOLESOM\ is planned for future work.
\begin{figure}
    \centering
    \includegraphics[width=1\linewidth,trim=0 21 0 0, clip]{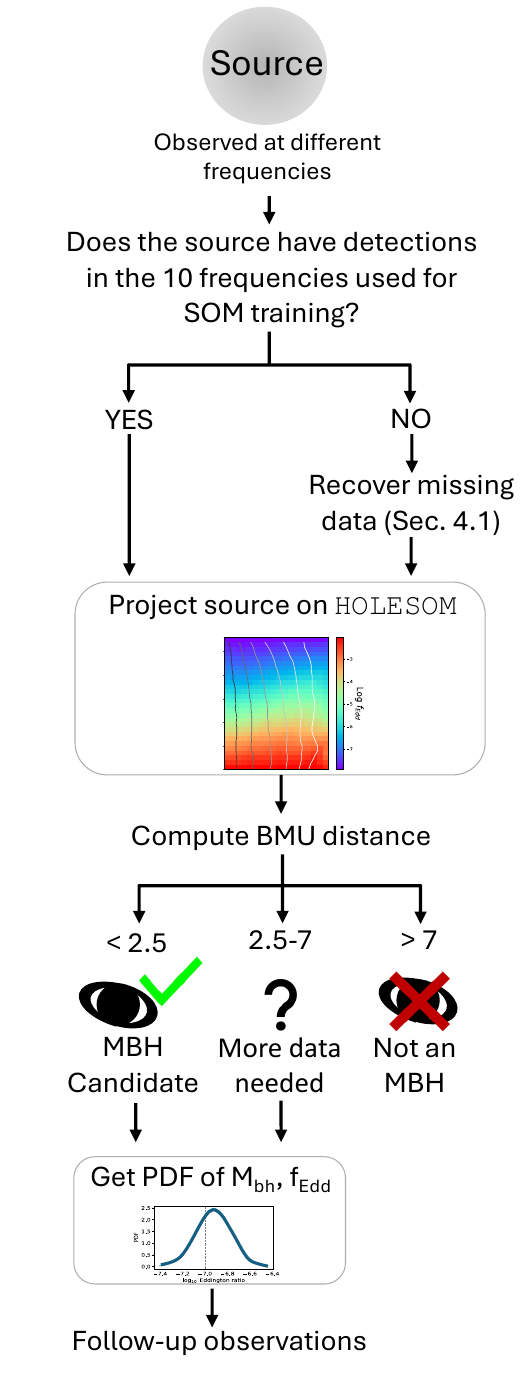}
    \caption{Schematic describing the procedure that \HOLESOM\ uses to identify MBH candidates and characterize their $\Mbh$ and $\fedd$ from multi-wavelength catalogs.}
    
    \label{fig:flowchart}
\end{figure}

In this section, we outline how the pre-trained SOM can be applied to photometric data of a new source, as schematically illustrated in Figure \ref{fig:flowchart}. The process involves projecting the available photometry onto the trained map using the same input features used during training. However, MBHs often have incomplete photometric coverage. In this case, we first recover missing observations at the required frequencies, as detailed in Section \ref{subsec:missing_data}. Now, with a full set of input features, the source is projected onto the map, and the BMU distance is computed. As detailed in Section \ref{subsec:exclusion_test}, based on this distance, the source is classified into one of three categories: i) high-confidence MBH candidate, ii) possible MBH candidate requiring additional observations for confirmation, and iii) not an MBH, likely belonging to a different astrophysical population. Once a candidate or possible candidate is identified, \HOLESOM\ provides an estimate, with uncertainties, of $\Mbh$ and $\fedd$, which could be refined with photometric or spectroscopic follow-up observations.

\section{Analytical Relations}
\label{sec:relations}
Unlike stellar-mass BHs, which follow reasonably tight relations between radio and X-ray luminosities (see, e.g., \citealt{Gallo_2003}), slowly-accreting MBHs do not exhibit such a strong correlation, as visually illustrated in Fig.\ref{fig:Lradio_X}. However, these luminosities depend on $\fedd$ and $\Mbh$. These insights drive us to seek analytical expressions tailored for slowly-accreting MBHs in a wide range of black hole masses and Eddington ratios. In the following, we present only those relations that effectively balance complexity and accuracy. 

\begin{figure}
\centering
    \includegraphics[width=0.5\textwidth,trim={1cm 0 0 0}]{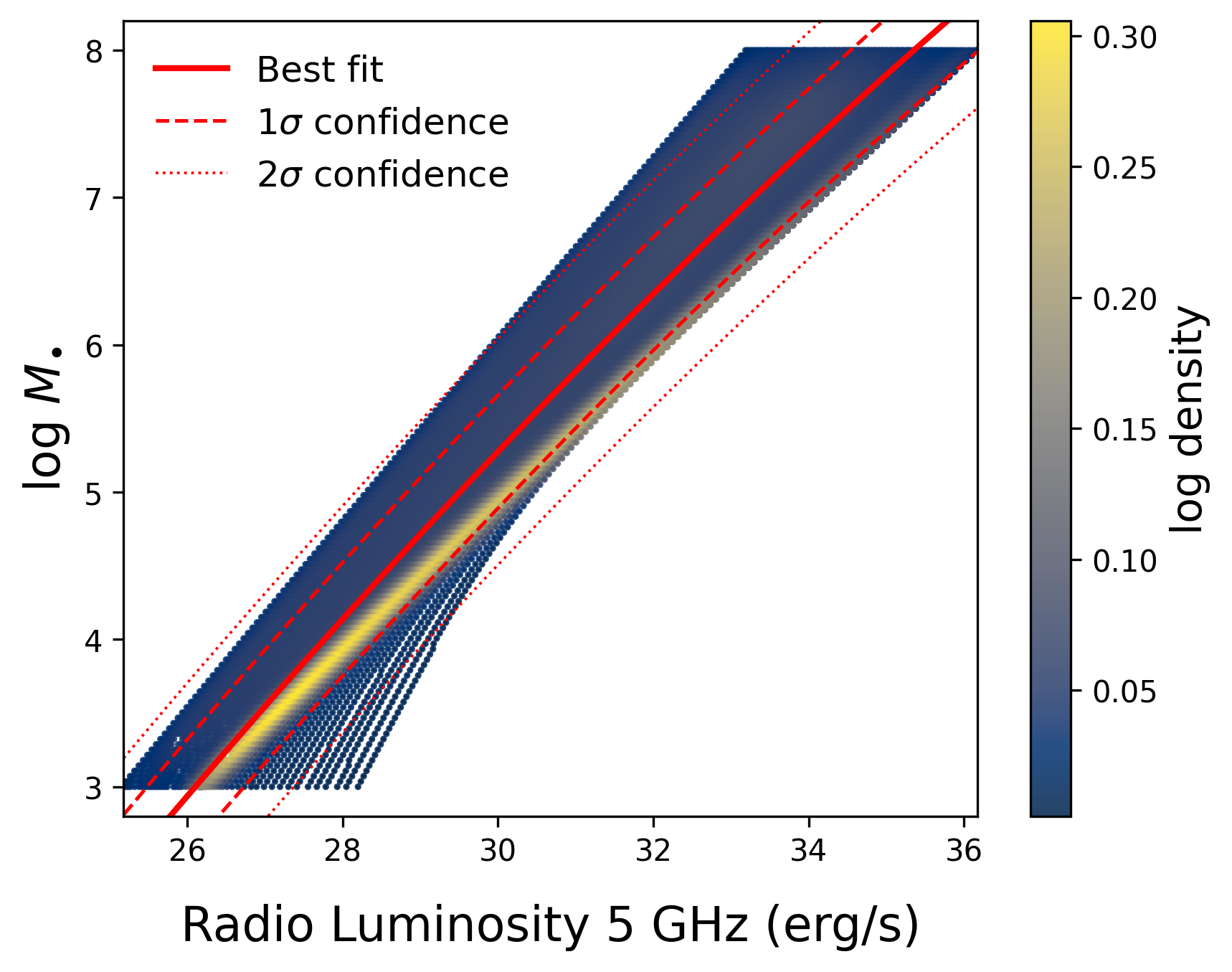}
    \includegraphics[width=0.5\textwidth,trim={1cm 0 0 0}]{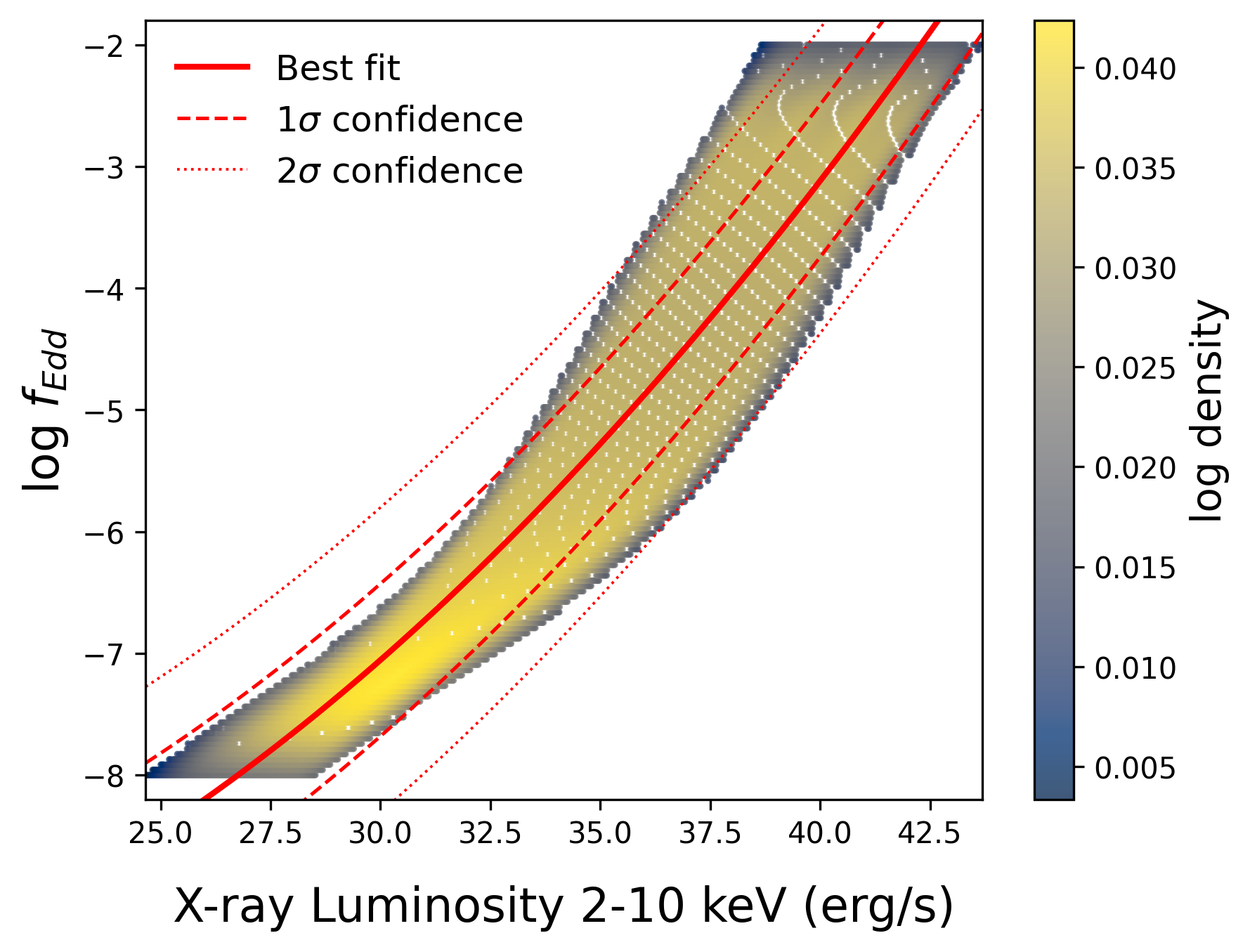}    
\caption{\textit{Top:} Black hole mass [$\rm M_{\odot}$] versus radio luminosity at 5 GHz, in units of erg/s. The red solid line represents the best fit of the distribution given by equation \ref{eq:lmass_lr}. The red dashed and dotted lines represent the 1$\sigma$ and 2$\sigma$ confidence levels. \textit{Bottom}: Eddington ratio versus luminosity in X-ray in erg$/$s. The red solid line and dashed and dotted lines are the same as in the top panel.}
\label{fig:lmass_lr}
\end{figure}

In the top panel of Figure \ref{fig:lmass_lr}, we present the relationship between $\Mbh$ and radio luminosity at 5 GHz, fitted using polynomial regression. The resulting equation is expressed as:
\begin{equation}
    \ell_m = \alpha_1 \ell_r^2 + \beta_1 \ell_r + \gamma_1 \, ,
    \label{eq:lmass_lr}
\end{equation}
where $\ell_m$ indicates Log($\Mbh$), $\ell_r$ is the logarithm of the radio luminosity at 5 GHz, $\alpha_1$, $\beta_1$, and $\gamma_1$ are constants whose values are reported in Table \ref{tab:constants}. 
In the the bottom panel of Fig.\ref{fig:lmass_lr}, we show the equivalent relation between $\fedd$ and the X-ray luminosity:
\begin{equation}
    \ell_f = \alpha_2 \ell_X^2 + \beta_2 \ell_X + \gamma_2 \, ,
    \label{eq:lfedd_lX}
\end{equation}
where $\ell_f$ indicates Log($\fedd$), $\ell_X$ is the logarithm of the integrated X-ray luminosity at 2-10KeV. The values of the constants $\alpha_2$, $\beta_2$, and $\gamma_2$ are also listed in Table \ref{tab:constants}.

\begin{figure}
\centering
    \includegraphics[width=0.5\textwidth]{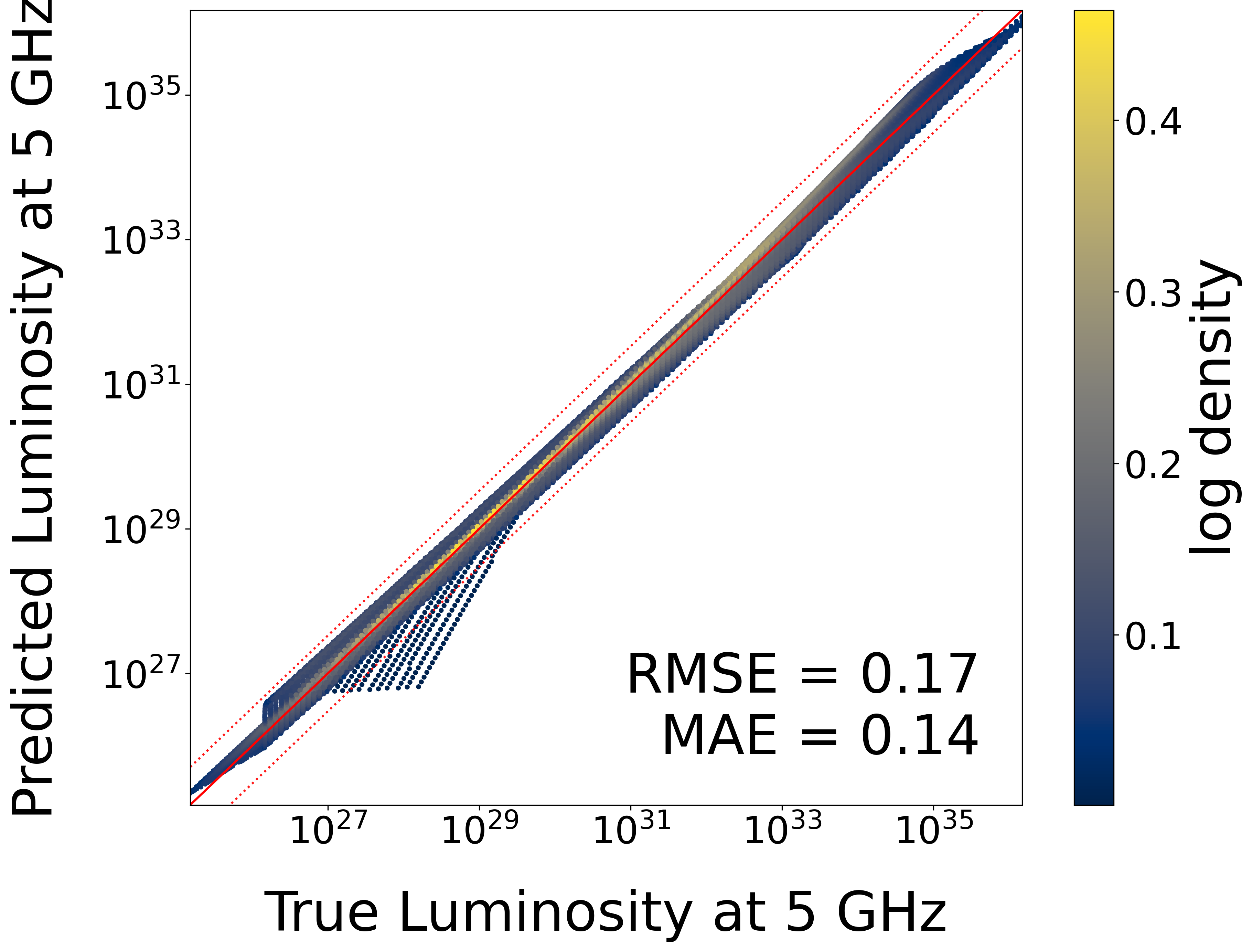}
\caption{Comparison between the true radio luminosity at 5 GHz and the predicted values as a function of X-ray luminosity and $\Mbh$, as inferred from Eq. \ref{eq:Lr_LX_mass}. The solid red line represents the 1:1 correlation, while the dotted lines indicate the 3$\sigma$ confidence level. The root mean square error and mean absolute error metrics are shown in the bottom-right corner.}
\label{fig:lr_lx_mass}
\end{figure}
\begin{table}[h!]
    \centering
    \caption{Constants used in the equations in Sec.\ref{sec:relations}.}
    \begin{tabular}{c c c}
        \midrule
        \textbf{$\Mbh = f(L_{5GHz})$} & \textbf{$\fedd = f(L_X)$} & \textbf{$L_{5GHz} = f(\Mbh, L_X)$} \\ 
         (eq.\,\ref{eq:lmass_lr}) &  (eq.\,\ref{eq:lfedd_lX}) & (eq.\,\ref{eq:Lr_LX_mass}) \\ 
         \midrule
         \midrule
        $\alpha_1$ = -0.0081 &  $\alpha_2$ = 0.0077 & $\alpha_3$ = 0.0220 \\ 
        $\beta_1$ = 1.0367 &  $\beta_2$ = -0.1442 & $\beta_3$ = 0.0404\\ 
          $\gamma_1$ =  -18.5593 & $\gamma_2$ = -9.6431 & $\gamma_3$ = -0.0220 \\ 
         & & $\delta_3$ = 1.0287 \\ 
        & & $\epsilon$ = 19.8177 \\ 
        \midrule
    \end{tabular}
    \label{tab:constants}
\end{table}

However, Figure \ref{fig:Lradio_X} shows that, at fixed X-ray luminosity, a source can exhibit either low or high radio luminosity, with the exception of very low and very high ends of the X-ray luminosity. This distribution is too complex for simple linear or polynomial models. Here, we turn to symbolic regression to capture $L_r = f(L_X, \fedd)$. 

The best equation from symbolic regression is: 
\begin{equation}
    \begin{aligned}
    \ell_r = & (\alpha_3 \ell_m + \beta_3) \ell_X \\
    & + (\gamma_3 \ell_m + \delta_3) \ell_m + \epsilon \, ,
    \end{aligned}
    \label{eq:Lr_LX_mass}
\end{equation}
with constants $\alpha_3$, $\beta_3$, $\gamma_3$, $\gamma_3$, and $\epsilon$ reported in Table \ref{tab:constants}.

As shown in Figure \ref{fig:lr_lx_mass}, the predicted radio luminosity closely matches the true values, except for a few points in the $10^{27}-10^{29}$ erg/s range. This is due to the abrupt decrease of $\Mbh$ at X-ray luminosity around $10^{38}-10^{39}$ erg/s, as seen in Fig.\,\ref{fig:Lradio_X}. In Fig.\,\ref{fig:lr_lx_mass}, we quantify the goodness of our prediction by computing the root mean square error (RMSE) and the mean absolute error (MAE).

\section{Conclusions}
\label{sec:conclusion}
Detecting MBHs accreting at extremely low rates in radiatively inefficient modes remains a significant challenge in astrophysics. These objects, which are easily confused with other classes of sources, require advanced techniques for accurate identification and characterization.

In this paper, we introduced \HOLESOM, a machine learning tool based on the SOM algorithm. Trained with $\sim20,000$ analytically-derived SEDs for ADAF accreting MBHs, \HOLESOM\: (i) automatically identifies and separates MBHs accreting in ADAF mode from other classes of objects, and (ii) efficiently characterize their physical properties, even with scarce data. We also used polynomial and symbolic regression to derive analytical relationships between radio/X-ray luminosities and black hole mass and Eddington ratio, providing additional insights into the behavior of MBHs radiating in ADAF mode.  

A key result of this work is \HOLESOM's ability to estimate mass and Eddington ratio ranges of MBHs with high accuracy, even in the presence of missing photometric data, which are recovered following \cite{LaTorre_2024}. As expected, the process becomes more challenging when the available data is scarce, particularly for MBHs accreting at very low Eddington ratios. In such cases, the accuracy of the mass and Eddington ratio estimates diminishes, but they are still consistent with their true values within uncertainties in the majority of cases. This highlights the importance of adding multi-wavelength data to improve the characterization of these objects. However, we demonstrated the robustness of our method by applying the model to the real-world case of \sag, where we retrieved distributions of $\Mbh$ and $\fedd$ that were in excellent agreement with known values. 

We also showed that a promising aspect of \HOLESOM\ is its ability to recognize the photometric signatures of MBHs while rejecting different astrophysical sources, based on their BMU distance. By projecting objects onto the trained SOM, MBHs exhibit low BMU distances ($<2.5$) and are placed in regions corresponding to their expected parameter values. In contrast, objects with significantly different SEDs, such as stellar-mass BHs, exhibit considerable high (i.e., $>$7) BMU distances and are identified as outliers, assigned to the edges of the map, or not projected onto the map at all. 

In conclusion, \HOLESOM\ provides a versatile and scalable framework for identifying and characterizing MBHs based on their SEDs. The capabilities of this method will become even more evident with large, multi-wavelength, low-z datasets. As part of our future work, we will apply \HOLESOM\ to a multi-wavelength, low-z catalog to identify slowly-accreting MBH candidates in the local Universe and provide preliminary estimates of their mass and Eddington ratio, constructing a first attempt at a mass and Eddington ratio distribution functions. Once MBH candidates are identified, additional multi-wavelength data will progressively hone in on the object's physical properties, optimizing the effectiveness of \HOLESOM\ and minimizing the impact of multiple missing photometric bands. As these surveys continue to grow, the ability to automatically identify MBHs and estimate their properties will be invaluable for advancing our understanding of the role these objects play in the evolution of galaxies.

\vspace{15pt}
\noindent \textit{Acknowledgments:} The authors are thankful to the anonymous referees for their thoughtful and constructive feedback, which greatly improved this manuscript. V.L.T also thanks Dr. Kaitlin Gili for insightful discussions on supervised and unsupervised methods during the writing of this paper.  V.L.T. acknowledges support from the Smithsonian Astrophysical Observatory Visiting Fellowship and the Kathryn A. McCarthy Graduate Fellowship in Physics administered by Tufts University. F.P. acknowledges support from a Clay Fellowship administered by the Smithsonian Astrophysical Observatory. This work was also supported by the Black Hole Initiative at Harvard University, funded by grants from the John Templeton Foundation and the Gordon and Betty Moore Foundation. This work used the FASRC cluster computer at Harvard University.

\vspace{10pt}
\noindent \textit{Software:} \textsc{SomPY} \citep{sompy}, \textsc{PySR} \citep{Cranmer_2023}, \texttt{numpy.polyfit}.


\bibliographystyle{aasjournal}

\appendix

\section{Ranges of $\fedd$ and $\Mbh$ in case of missing data}
\label{app:missing_data}
\begin{figure*}[ht!]
    \centering
    \includegraphics[width=0.4\textwidth]{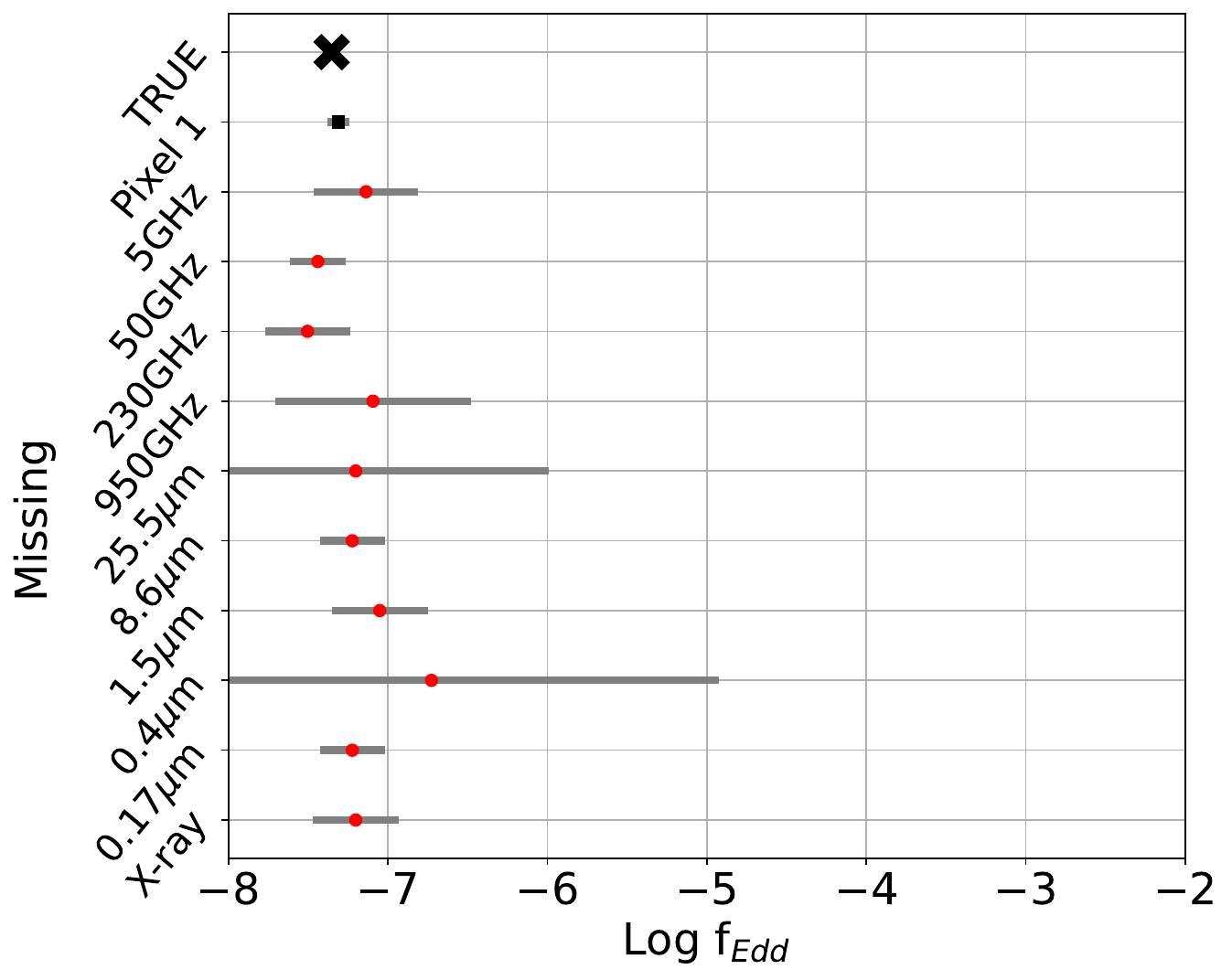}\includegraphics[width=0.4\textwidth]{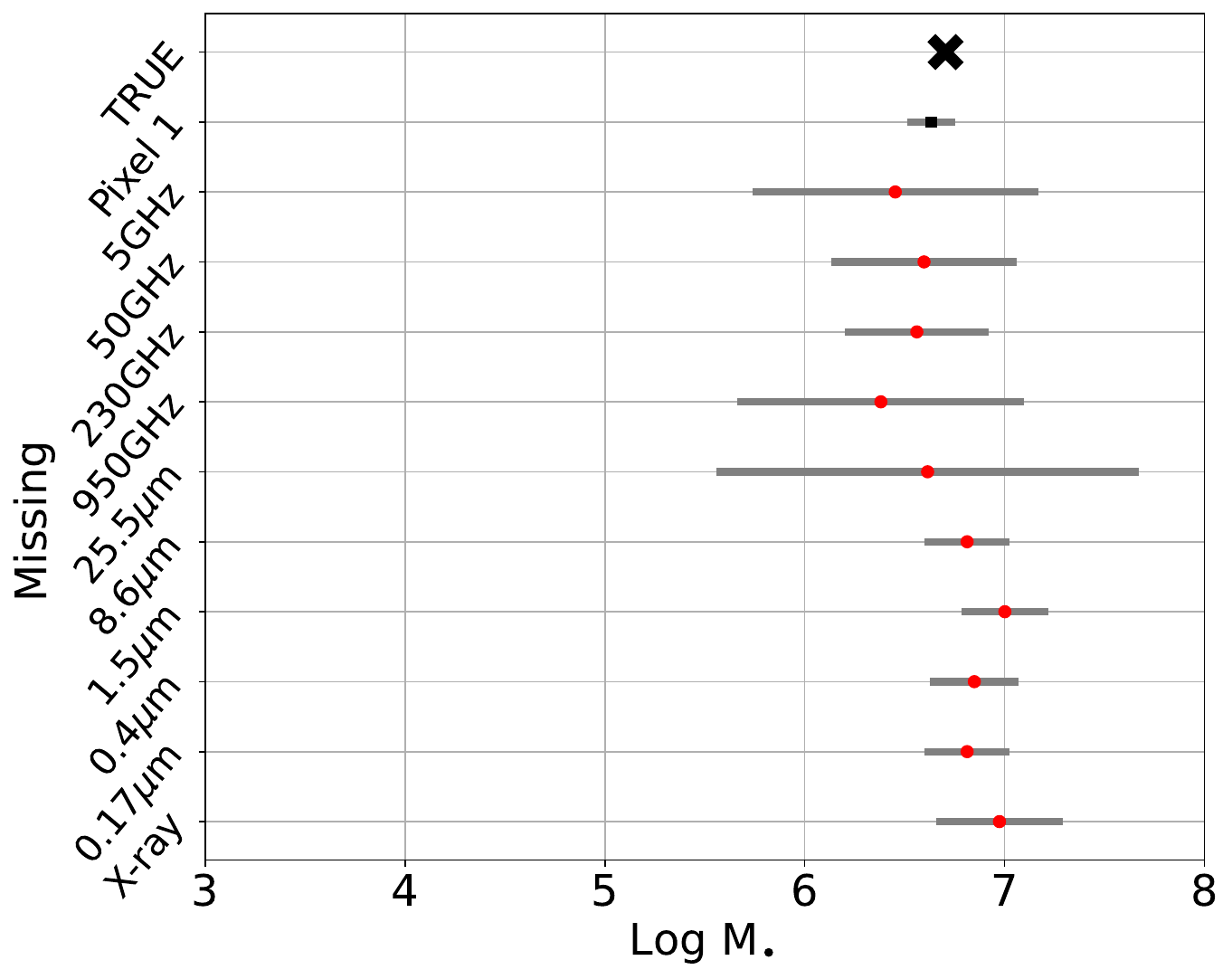}
    
    \includegraphics[width=0.4\textwidth]{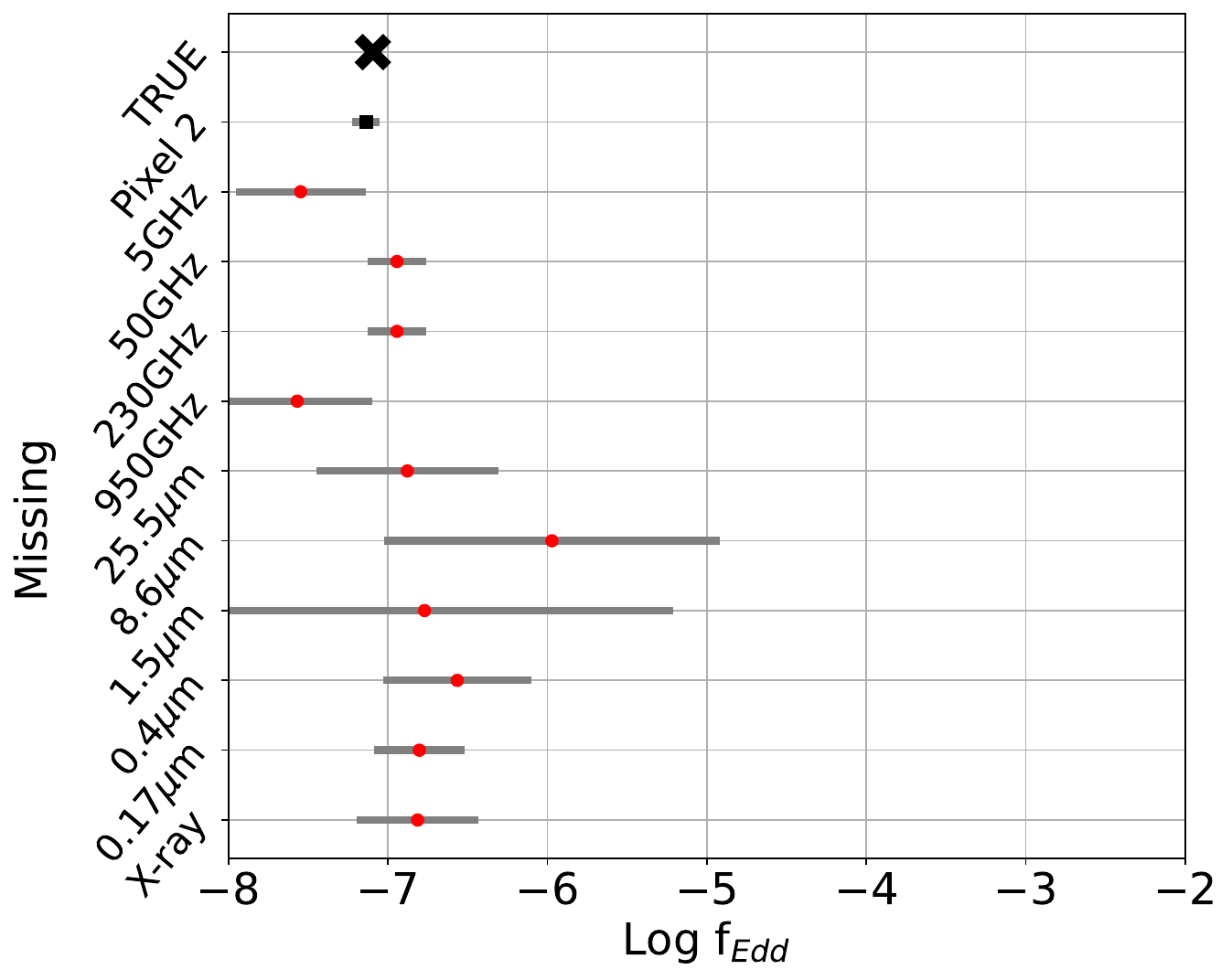}\includegraphics[width=0.4\textwidth]{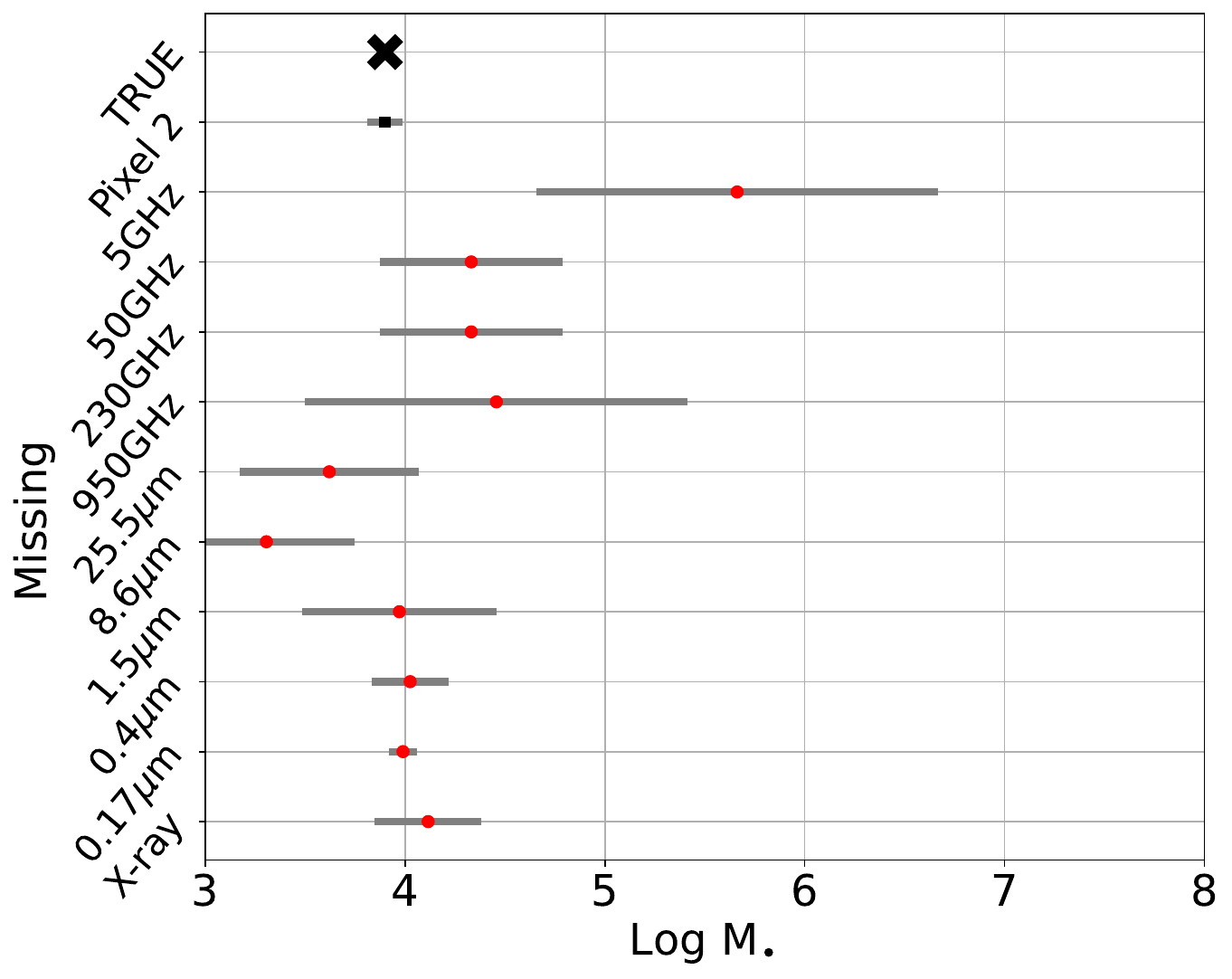}

    \includegraphics[width=0.4\textwidth]{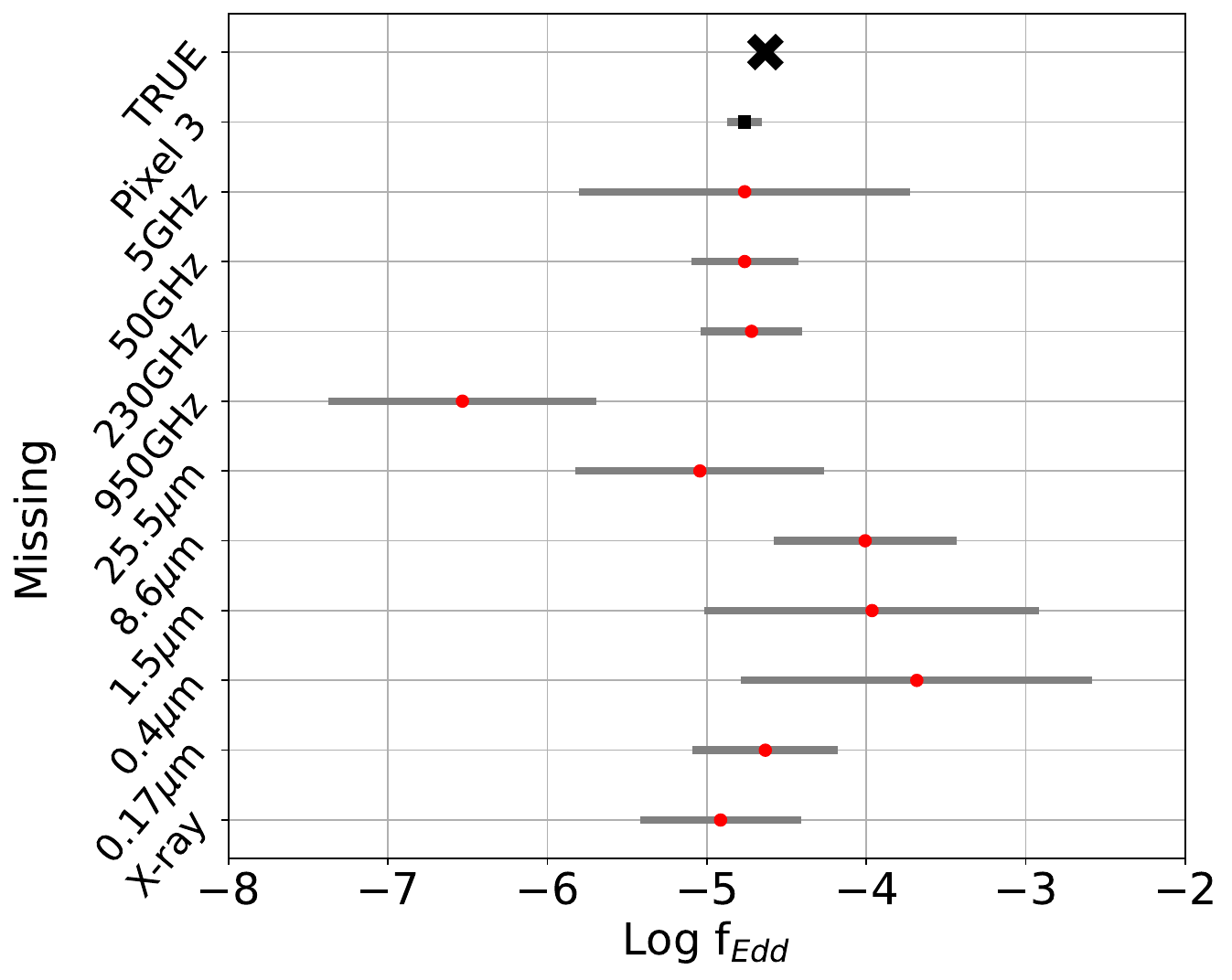}\includegraphics[width=0.4\textwidth]{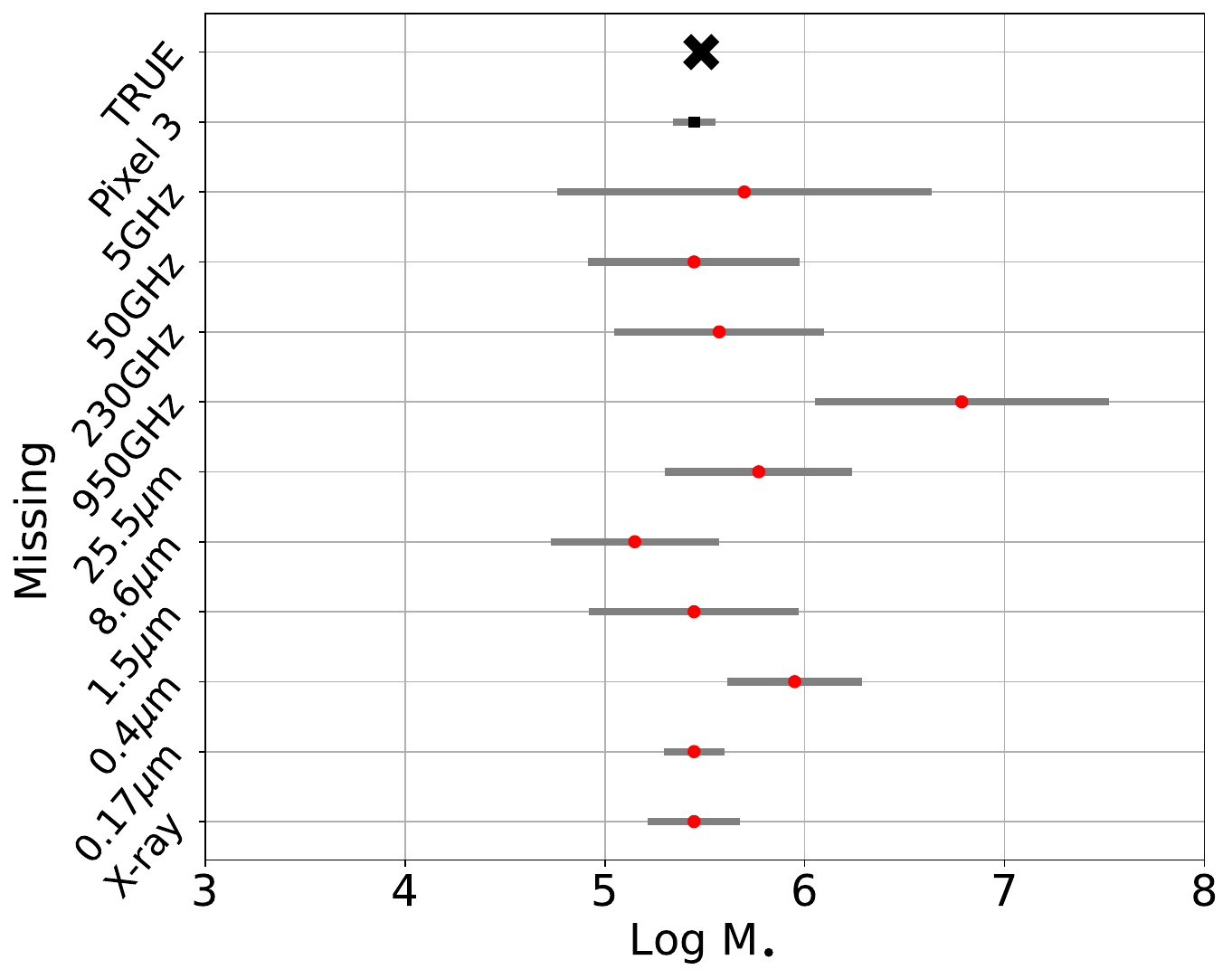}
    
    \caption{\textit{Left panels}: Ranges of Log $\fedd$ when a missing observation is recovered. The red circle represents the median of the distribution of possible Log $\fedd$ values once the missing observation is recovered. The grey line covers the standard deviation of that distribution. The missing observations are listed on the y-axis. The label `TRUE', indicated by a black cross, refers to the true value of the test BH residing in one of the five test pixels. The latter are listed on the second label on the y-axis. The median of the distribution of Log $\fedd$ in that pixel is indicated by a black square, and the grey line represents the standard deviation of that distribution in that pixel. 
    \textit{Right panels}: Same as the left panels, but for $\log_{10}\Mbh$ [$\Msun$]. }
    \label{fig:results}
\end{figure*}

\begin{figure*}
    \centering
   
    \includegraphics[width=0.4\textwidth]{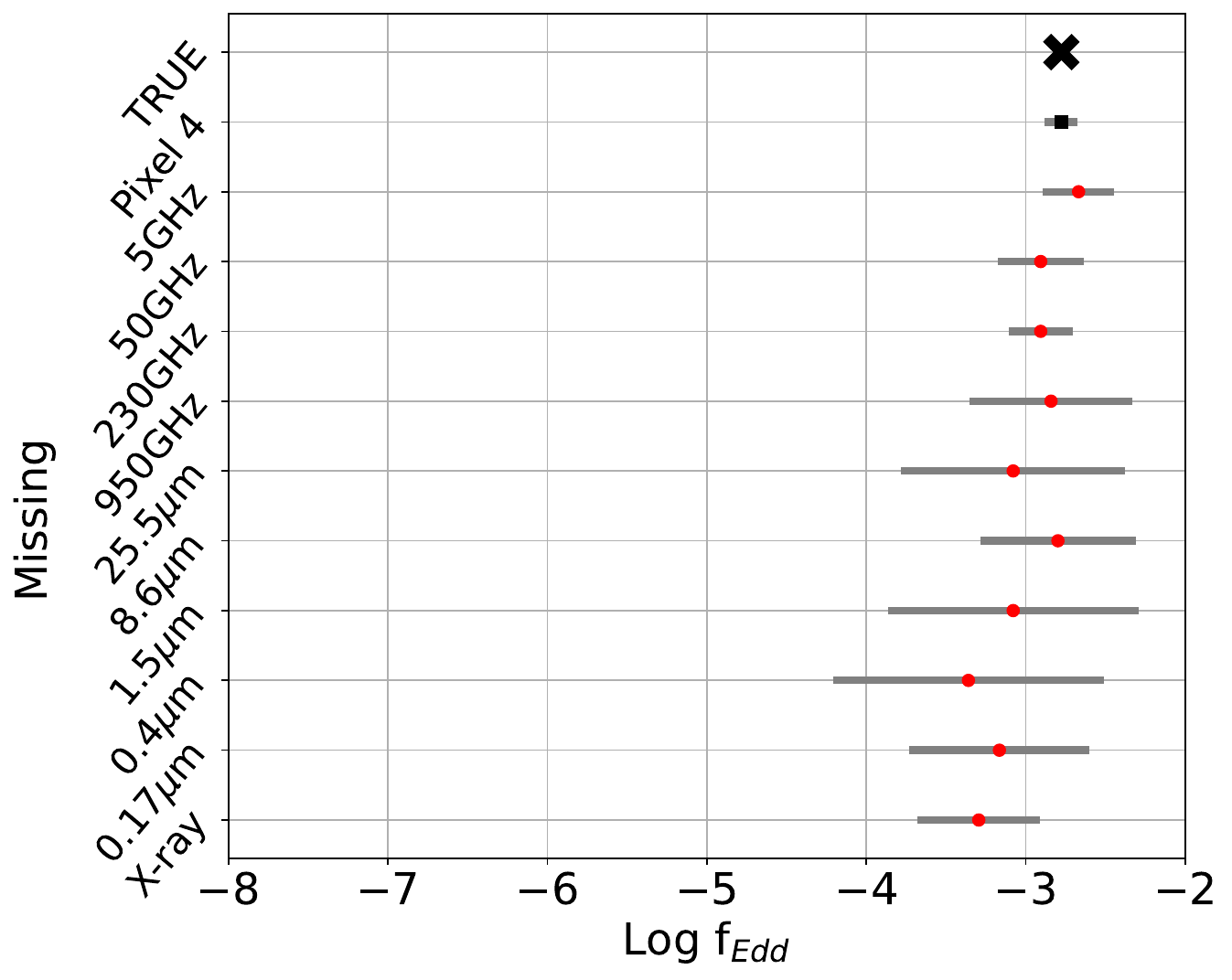}\includegraphics[width=0.4\textwidth]{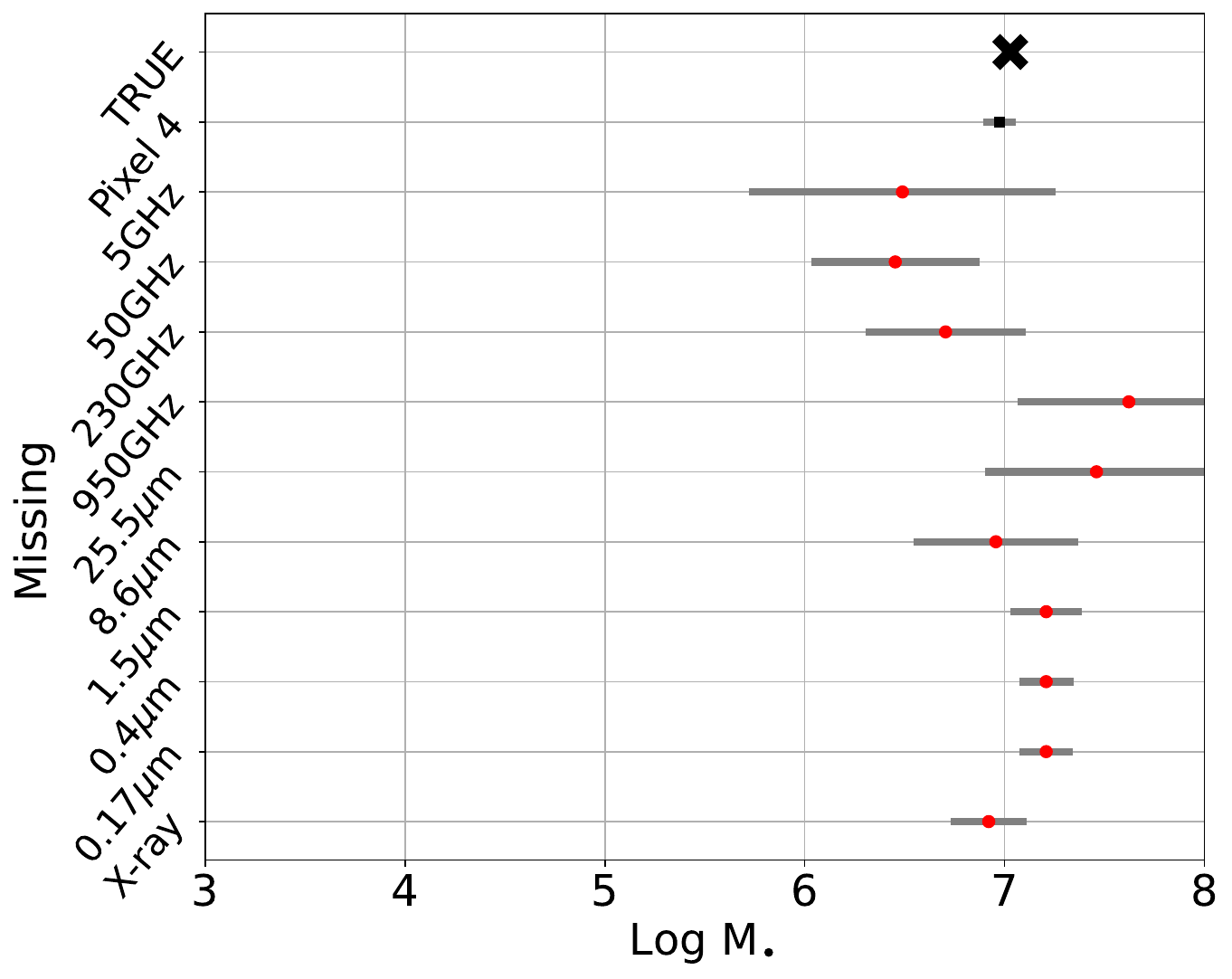}

    \includegraphics[width=0.4\textwidth]{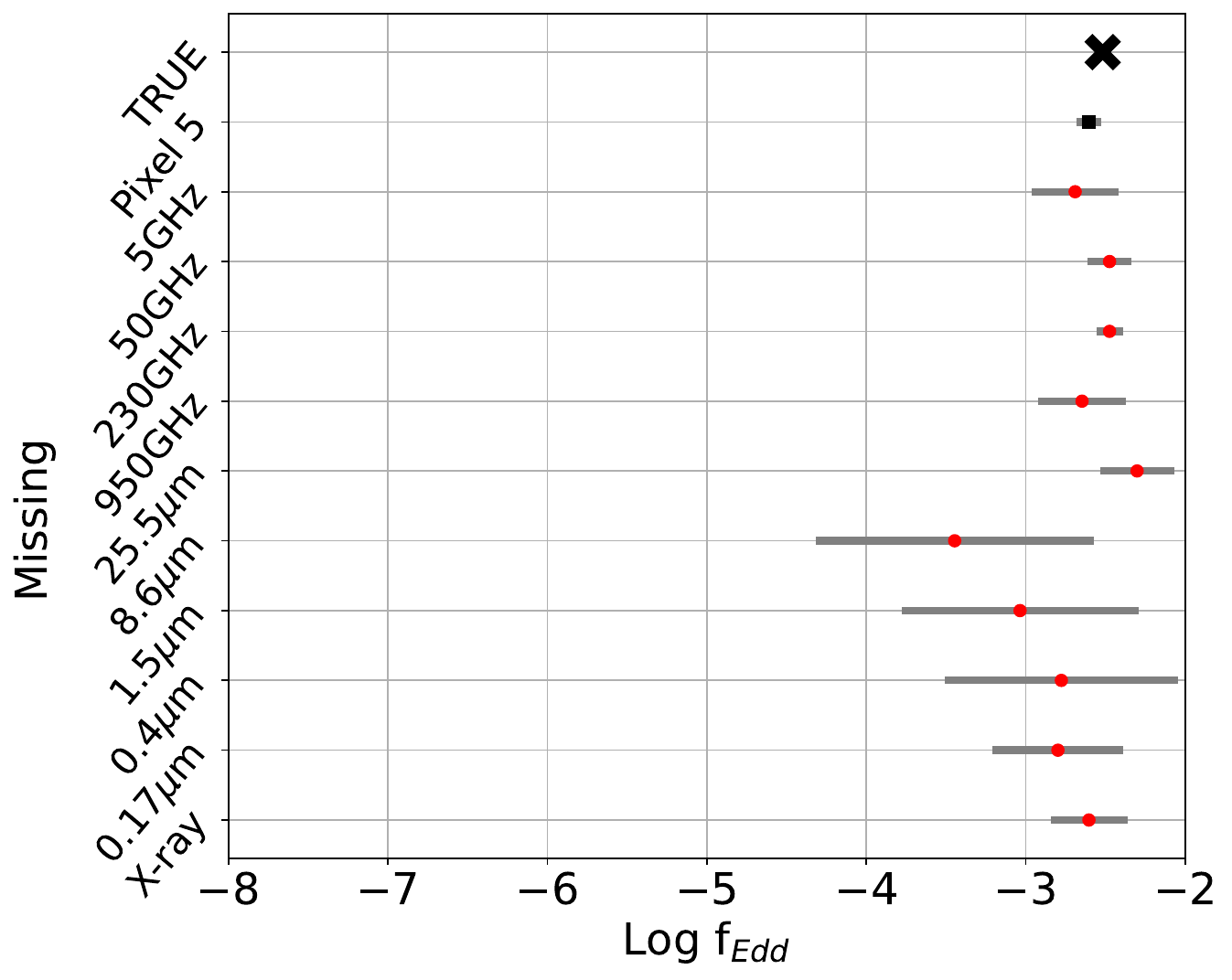}\includegraphics[width=0.4\textwidth]{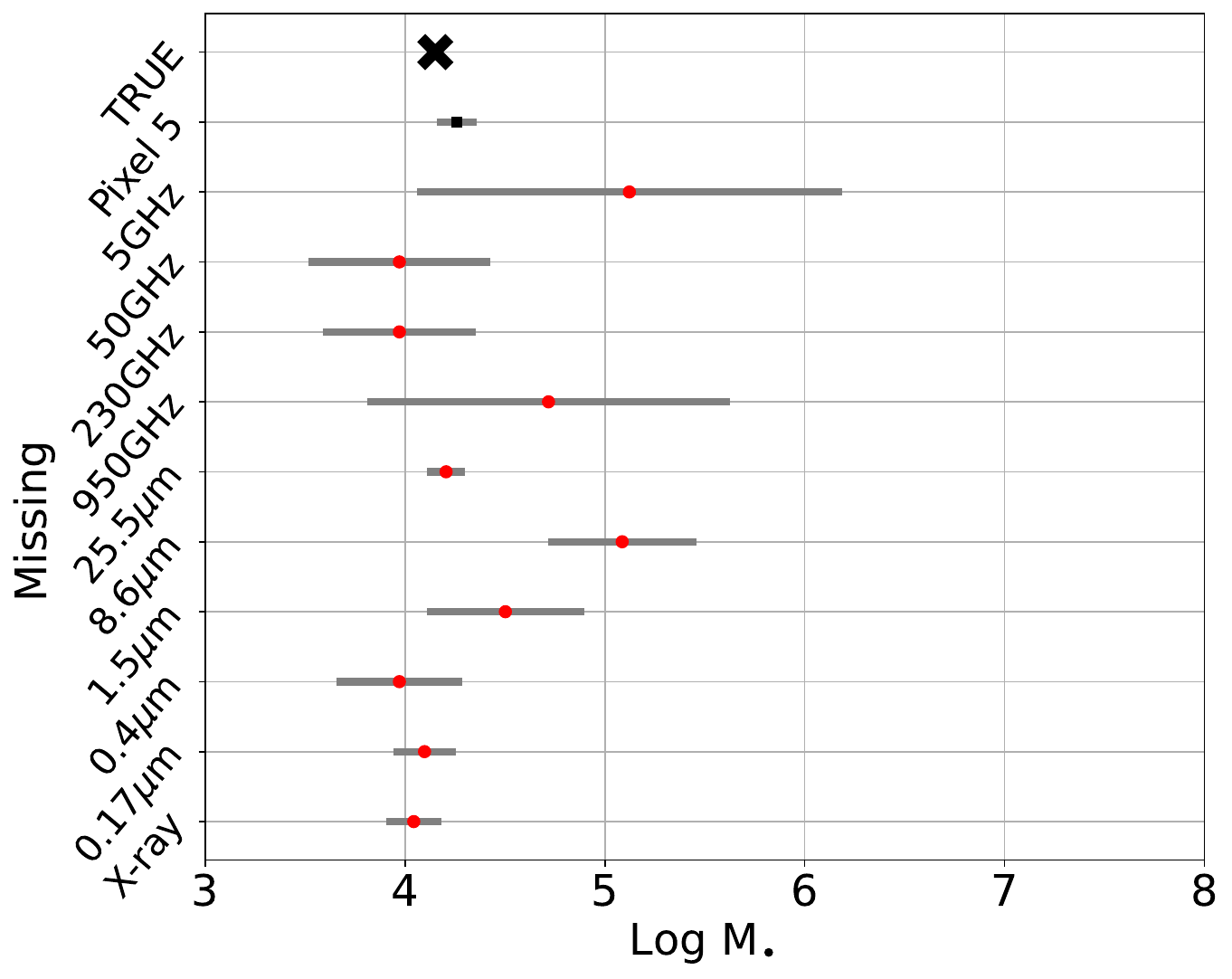}
    
    \caption{Same as Figure\,\ref{fig:results} but for test MBHs in pixels 4 and 5. }
    \label{fig:results1}
\end{figure*}

\begin{figure}
    \centering
    \includegraphics[angle=90,origin=c,width=0.5\textwidth]{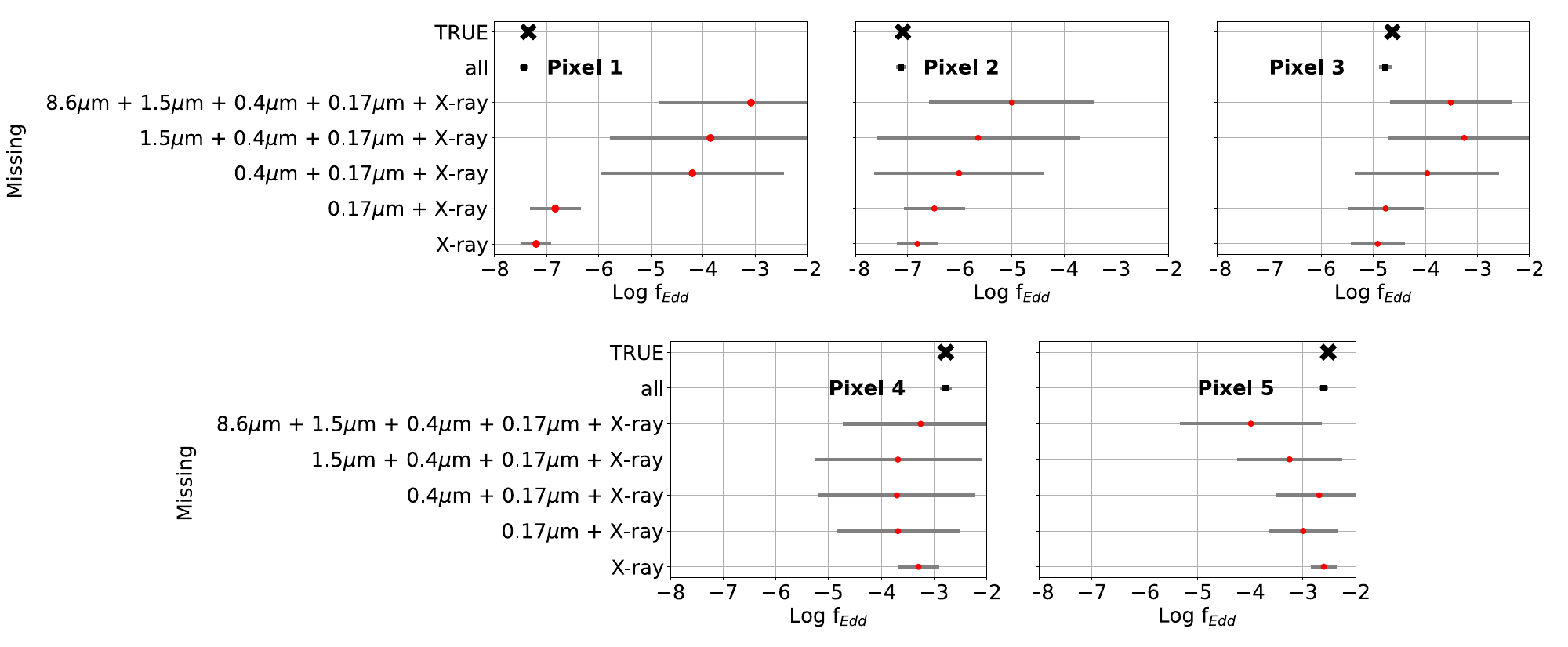}
    
    \caption{Range of Log $\fedd$ for test MBHs from the five test pixels (indicated in bold text) in the case of recovered missing multiple observations, as listed on the y-axis. The black cross and square for the cases 'TRUE' and 'all' are the same as in Figures\,\ref{fig:results}-\ref{fig:results1}.}
    \label{fig:fedd_more_missing}
\end{figure}

\begin{figure}
    \centering
    \includegraphics[angle=90,origin=c,width=0.5\textwidth]{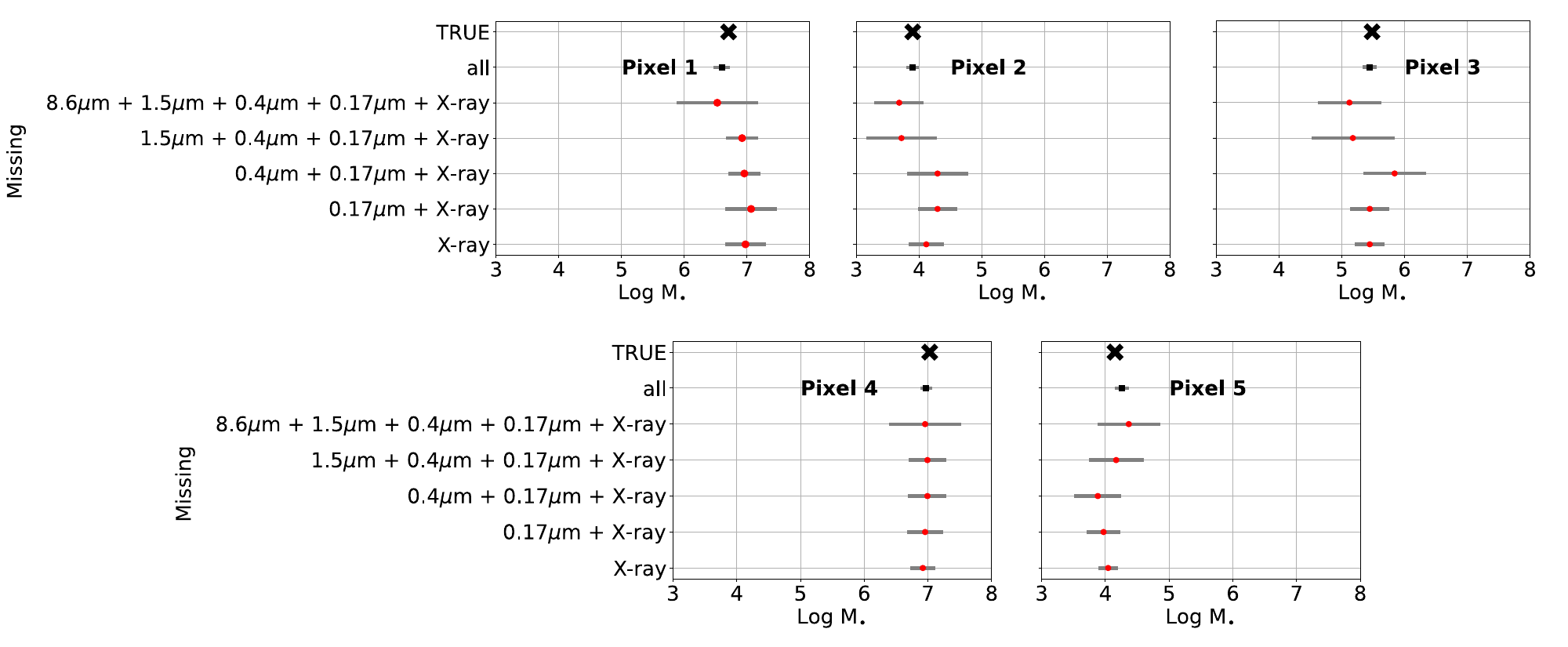}
    \caption{Same as in Figure\,\ref{fig:fedd_more_missing} but for Log $\Mbh$ [$\Msun$].}
    \label{fig:mass_more_missing}
\end{figure}

\end{document}